\documentclass[iop,apj]{emulateapj}
\usepackage{amsmath}

\newcommand\msol{{\cal M_{\odot}}}
\newcommand\teff{{T_{\rm eff}}}

\newcommand\pab{$\langle P_{ab}\rangle$}
\newcommand\pc{$\langle P_c\rangle$}
\newcommand\per{$\langle P\rangle$}
\newcommand\tavg{$\langle \log\,T_{\rm eff}\rangle$}
\newcommand\mavg{$\langle {\cal M}/{{\cal M}_\odot}\rangle$}
\newcommand\bavg{$\langle M_{\rm bol}\rangle$}
\newcommand\vavg{$\langle M_V\rangle$}
\newcommand\delv{$\Delta V_{TO}^{HB}$} 
\newcommand\amlt{\alpha_{\rm MLT}}
\newcommand\lta{\mathrel{\hbox{\raise 0.6 ex \hbox{$<$}\kern
                   -1.8 ex\lower .5 ex\hbox{$\sim$}}}}
\newcommand\gta{\mathrel{\hbox{\raise 0.6 ex \hbox{$>$}\kern
                   -1.7 ex\lower .5 ex\hbox{$\sim$}}}}

\begin{document}

\epsscale{1.1}

\title{CONSTRAINTS ON THE DISTANCE MODULI, HELIUM AND METAL ABUNDANCES, AND
AGES OF GLOBULAR CLUSTERS FROM THEIR RR LYRAE AND NON-VARIABLE HORIZONTAL-BRANCH
STARS.~I.~M\,3, M\,15, AND M\,92}

\shortauthors{VandenBerg, Denissenkov, \& Catelan}
\shorttitle{HB Stars in M\,3, M\,15, and M\,92}

\author{
Don A.~VandenBerg\altaffilmark{1}, 
P.~A.~Denissenkov\altaffilmark{1}, 
and
M\'arcio Catelan\altaffilmark{2,3}
}

\altaffiltext{1}{Department of Physics \& Astronomy, University of Victoria, 
P.O.~Box 1700, STN CSC, Victoria, BC, V8W 2Y2, Canada; vandenbe@uvic.ca,
pavelden@uvic.ca}

\altaffiltext{2}{Pontificia Universidad Cat\'olica de Chile, Instituto de
Astrof\'isica, Facultad de F\'isica, Av.~Vicu\~na Mackena 4860, 782-0436 Macul,
Santiago, Chile; mcatelan@astro.puc.cl}

\altaffiltext{3}{Millennium Institute of Astrophysics, Santiago, Chile}


\submitted{Submitted to The Astrophysical Journal}

\begin{abstract}

Up-to-date isochrones, zero-age horizontal branch (ZAHB) loci, and evolutionary
tracks for core He-burning stars are applied to the color-magnitude diagrams
(CMDs) of M\,3, M\,15, and M\,92, focusing in particular on their RR Lyrae
populations.  Periods for the $ab$- and $c$-type variables are calculated using
the latest theoretical calibrations of $\log\,P_{ab}$ and $\log\,P_c$ as a
function of luminosity, mass, effective temperature ($\teff$), and metallicity.
Our models are generally able to reproduce the measured periods to well
within the uncertainties implied by the stellar properties on which pulsation
periods depend, as well as the mean periods and cluster-to-cluster differences in
\pab\ and \pc, on the assumption of well-supported values of $E(B-V)$,
$(m-M)_V$, and [Fe/H].  While many of RR Lyrae in M\,3 lie close to the same
ZAHB that fits the faintest HB stars at bluer or redder colors, the M\,92
variables are all significantly evolved stars from ZAHB locations on the blue
side of the instability strip.  M\,15 appears to contain a similar population
of HB stars as M\,92, along with additional helium-enhanced populations not
present in the latter which comprise most of its RR Lyrae stars.  The large
number of variables in M\,15 and the similarity of the observed values of
\pab\ and \pc\ in M\,15 and M\,92 can be explained by HB models that allow for
variations in $Y$.  Similar ages ($\sim 12.5$ Gyr) are found for all three
clusters, making them significantly younger than the field halo subgiant
HD\,140283.  Our analysis suggests a preference for stellar models that take
diffusive processes into account.

\end{abstract}


\section{Introduction}
\label{sec:intro}

The globular clusters (GCs) M\,15 (NGC\,7078) and M\,92 (NGC\,6341) are
generally thought to have very close to the same metallicity (see the
spectroscopic surveys by e.g., \citealt{ki03}, \citealt{cbg09a}) and age (e.g.,
\citealt{san82}, \citealt{van00}, \citealt{dsa10}).  The strongest argument in
support of coevality is that color-magnitude diagram (CMD) studies have shown
that the difference in magnitude between the turnoff (TO) and the horizontal
branch (HB) is nearly identical for these two systems (e.g., \citealt{dh93}). 
Originally, this so-called ``\delv\ parameter" was measured at the color of the
TO (\citealt{ir84}), but the uncertainty of $V_{TO}$ can easily be as high as
$\pm 0.1$ mag, implying $\delta$(age) $\sim\pm 1$ Gyr, because of the difficulty
of determining the magnitude of the bluest point in a sequence of stars that is,
by definition, vertical at the TO.  Much more precise ages can be derived by
fitting isochrones to the arc of stars from $\sim 1$ mag below the TO through to
a point on the subgiant branch (SGB) that is $\sim 0.05$ mag redder than the TO,
in conjunction with fits of zero-age horizontal branch (ZAHB) models to the
cluster HB populations (\citealt[hereafter V13]{vbl13}; \citealt{lvm13}).  Using
this technique, which builds on the approaches advocated by \citet{cdk96} and
\citet{bcp98}, V13 found that M\,15 and M\,92 have the same age to within $\pm
0.25$ Gyr.  [The shapes of modern isochrones in the vicinity of the TO, in
particular, appear to be quite a robust prediction and, in fact, stellar models
are able to reproduce the turnoff portions of observed CMDs rather well when
up-to-date color--$\teff$ relations (e.g., \citealt[hereafter CV14]{cv14}) are
employed; see V13.]

However, this result is not yet ironclad --- primarily because the two GCs have
very different HB morphologies.  Indeed, M\,15 is not at all like the majority
of clusters with [Fe/H] $< -2$, including M\,92, whose HB populations are
located predominately to the blue of the instability strip (IS), and their RR
Lyrae stars constitute just a small fraction of the total number of core 
helium-burning stars.  In M\,92-like HBs, both the paucity of variables and
their high pulsation periods, relative to those determined for RR Lyrae in more
metal-rich clusters (like M\,3), can be plausibly explained if these variables
evolved into the IS from ZAHB locations on the blue side of the IS, where most
of the HB stars are found (\citealt{ren83}, \citealt{ldz90}, \citealt{psc02},
\citealt{scf14}).  

Curiously, M\,15 has a horizontal branch that spans a much
wider range in color than is typical of extremely metal-deficient GCs, and it
is so rich in RR Lyrae that a large fraction of its variables must have
evolved from ZAHB structures inside the IS (\citealt{rtr84}, \citealt{bcd84},
\citealt{bcf85}, \citealt{rf88}).  Yet, the mean period of its $ab$-type RR
Lyrae stars agrees very well with the values of \pab\ that have been derived
for other Oosterhoff type II (hereafter, Oo II) systems
(\citealt[\citealt{oo44}]{oo39}), including M\,92; see \citet[his Table
2]{cat09}.  This suggests that, at the same intrinsic color, M\,15 and M\,92
variables have similar luminosities; and therefore that (in the mean at least)
M\,15 RR Lyrae lie above the extension into the IS of the same ZAHB which
provides a good fit to the main non-variable, blue HB population of M\,92 (as
well as its counterpart in M\,15).   

There is another important difference between M\,15 and other GCs of very low
metallicity in that it is the only one which has been found to have a signficant
dispersion in the abundances of heavy neutron-capture elements (e.g.,
\citealt{coh11}, \citealt{whs13}).  This may be (probably is) connected with
the fact that M\,15 is one of the most luminous, and thus most massive,
clusters in the Galaxy.  Indeed, other systems with integrated $M_V < -9$ (see
the latest version of the \citealt{har96}
catalogue\footnote{www.physics.mcmaster.ca/$\sim$harris/mwgc.dat}) generally
exhibit the largest chemical abundance anomalies; see, for instance, recent
investigations of 47 Tuc (\citealt{mpd12}, \citealt{gls13}), NGC\,2808
(\citealt{car15}, \citealt{mmp15}), NGC\,2419 (\citealt{ck12}, \citealt{mbi12}),
NGC\,6441 (\citealt{bpm13}) and M\,2 (\citealt{yrg14}).  

Moreover, as discussed in, e.g., the studies of 47 Tuc by \citet{dvd10} and of
NGC\,2808 by \citet{dsf11} and \citet{mmp14}, consequences of the observed (or
inferred, in the case of helium) abundance variations for their HB populations
can often be identified.  To be specific, Di Criscienzo et al.~found that
the best match to the observed HB morphology of 47 Tuc is obtained if synthetic
HB populations are generated on the assumption of $\Delta Y \approx 0.03$ for
the initial He abundances (also see \citealt{scp16}).  This is approximately
the dispersion in $Y$ that has been inferred from the width of the cluster MS
by \citet{apk09}.  Similarly, D'Alessandro et al.~and Marino et al.~have found
that the very unusual HB of NGC\,2808 can be explained if it consists of
sub-populations of stars with low, intermediate, and high helium abundances
that are consistent with the values of $Y$ implied by the cluster's triple MS
(see \citealt{pba07}).  Hence, it may turn out that the HB of M\,15 cannot be
satisfactorily explained except as a superposition of multiple stellar
populations --- something which has long been suspected (see, e.g.,
\citealt{bcf85}).  

Indeed, \citet[also see \citealt{jl15}]{jlj14} have recently speculated that the
presence of different {\it generations} of stars, which assumes that resident
chemically distinct populations formed at different times (\citealt{gcb12}),
may be responsible for the appearance of the observed HBs in {\it most}
clusters, as well as their separation into Oosterhoff groups.  In their
scenario, core helium-burning stars with normal helium abundances ($Y \approx
0.25$) populate a different range in color on the HB than those with slightly
higher $Y$, enhanced CNO abundances, and younger ages (by 1--2 Gyr), and (if
they exist) still younger stars with much higher $Y$.  That is, the spread in
color on the HB would be due more to the differences in the ages and the
abundances of helium and CNO of the existing subgroups than to a large
dispersion in mass at nearly constant $Y$ and [CNO/Fe], which is the canonical
explanation (\citealt{ro73}).  Since HBs are shifted to the red as the
metallicity increases, the stars that are located in the IS could belong mostly
to the first, second, or third generation depending on the cluster [Fe/H],
possibly producing the observed RR Lyrae period shifts (see Jang et al., their
Fig.~1 and the accompanying discussion.).  

However, although difficult to measure, C$+$N$+$O appears to be constant to
within measuring uncertainties in most GCs; see, e.g., the spectroscopic results 
obtained for M\,4 by \citet{sci05}, for NGC\,6397 and NGC\,6752 by 
\citet{cgl05}, and for M\,3 and M\,13 by \citet[also see \citealt{ssb96}]{cm05}.
To date, there is compelling evidence for large star-to-star [CNO/Fe] variations
only in NGC\,1851 (\citealt{ygd09}), though there is some suggestion from
photometric data that 47 Tucanae harbors a minor CNO-enhanced population of
stars in its core (\citealt{apk09}).  As shown by \citet{csp08} in the case of
NGC\,1851, large variations in [CNO/Fe] cause the SGB to be broadened, or split,
and since this is not commonly seen in GC CMDs (see, e.g., the {\it HST}
photometric survey carried out by \citealt{sbc07}), intrinsic spreads in
[CNO/Fe] larger than $\sim 0.2$ dex are effectively ruled out (unless the
effects of age and [CNO/Fe] variations compensate each other).  Indeed, even
well-developed O--Na and Mg--Al anti-correlations, such as those derived for
stars in M\,13 by \citet{jp12} and \citet{dny13}, respectively, can be
reproduced remarkably well by theoretical models if the H-burning occurs at a
sufficiently high temperature ($\approx 75\times 10^6$ K) and both C$+$N$+$O
and the total number of Mg and Al nuclei are constant (see \citealt{dvh15}).

At the present time, supermassive stars (\citealt{dh14}) are the only known
nucleosynthesis site that has the required H-burning temperatures to achieve
this consistency between theory and observations without requiring large
{\it ad hoc} modifications to the rates of relevant nuclear 
reactions.\footnote{\citet{rdc15} have pointed out some difficulties with the
scenario proposed by \citet{dh14}, and we do not disagree that there are valid
concerns (also see \citealt{ikp16}).  However, they may simply be telling us
that we do not yet have the
correct understanding of how supermassive stars fit into our picture of the very
early evolution of GCs, or whether they are but one of the contributors to the
chemical evolution of these systems at early times.  Given their considerable
success in explaining the observed light-element abundance correlations and
anti-correlations --- and the limited success, or outright failure, of other
hypotheses to accomplish the same thing (see \citealt{dvh15}) --- we suspect
that supermassive stars will turn out to be an important piece of the puzzle.
Although it is commonly believed that the chemically distinct populations in
GCs arose as a result of successive star formation events, this possibility is
still conjecture at the present time.  The CN-poor, O-rich, Na-poor, $\ldots$
stars could have formed at essentially the same time as the CN-rich, O-poor,
Na-rich, $\ldots$\ stars if such chemical abundance variations within GCs have,
e.g., a supermassive star origin.}
Thus there are ample reasons to question the variations
in CNO and age that underpin the explanation of the Oosterhoff dichotomy
suggested by \citet{jlj14} and \citet{jl15}.  To properly evaluate the validity
of their proposals, one should first examine how well updated models for the
evolution of HB stars are able to explain both the morphologies of the observed
HBs in GCs and the periods of their RR Lyrae variables.  Since the difference
in \pab\ between Oo II systems (M\,15, M\,92) and Oo I clusters (e.g., M\,3) is
of particular interest, a careful consideration of the M\,3 HB is included in
this investigation. 

After describing our evolutionary computations in \S\ref{sec:models}, fits of
isochrones to the cluster TOs and of evolutionary tracks for the core He-burning
phase to the observed HBs are presented in \S\ref{sec:obs}, along with
comparisons of the predicted and observed periods of their RR Lyrae.  The main
results of this study are summarized and briefly discussed in \S\ref{sec:sum}.

\section{Stellar Evolutionary Models}
\label{sec:models}

All stellar models that are used in this investigation to fit the main sequence
(MS) and red giant branch (RGB) photometric sequences of GCs were generated
using the Victoria evolutionary code, as described in considerable detail by
\citet{vbd12}.  To be specific, we have made use of the computations for
[$\alpha$/Fe] $= +0.4$ from \citet[hereafter V14]{vbf14}, since this is
approximately the observed enhancement of the $\alpha$-elements in metal-poor
clusters (e.g., \citealt{cbg09b}), as well as several new grids that allow for
[O/Fe] values as high as $+1.0$ (i.e., $0.0 \le \delta$[O/Fe] $\le 0.6$
above the amount implied
by the adopted value of [$\alpha$/Fe]).  (The latter represent just a small
subset of the extensive sets of tracks and isochrones, to be made publicly
available in a forthcoming paper, in which [O/Fe] is treated as a free
parameter.)  Both M\,92 and M\,15, in particular, could be expected to have
high oxygen abundances if the variation of [O/Fe] with [Fe/H] that has been
derived for extremely metal-deficient stars in the Galaxy (\citealt{dkb15},
\citealt{aac15}) applies to them.  In fact, it may not be possible to explain
the reddest HB stars in M\,15 without assuming very high oxygen abundances
(see \S\ref{subsec:m15}).  As documented in the Appendix of the paper by
V14, the elegant interpolation software developed by P.~Bergbusch
enables us to generate isochrones for arbitrary [Fe/H], $Y$, and [O/Fe] within
the ranges for which evolutionary tracks have been computed.

Because a suitable treatment of semi-convection or core overshooting in
helium-burning stars has not yet been incorporated into the Victoria code,
the evolution of stars past their ZAHB locations has never been followed.
However, it has already been demonstrated (see \citealt{vbd12}) that tracks for
the MS and RGB phases are nearly identical with those predicted by the MESA
code (\citealt{pbd11}) when very similar physics is assumed.  If similar good
agreement is found in the case of the respective ZAHB models, then no
significant inconsistencies are introduced by using the MESA code to generate
ZAHB loci and post-ZAHB tracks while employing Victoria isochrones to describe
the earlier evolutionary phases.  [The main advantage of this approach is that
the Victoria code contains an implementation of the \citet{egg71} non-Lagrangian
method of solving the stellar structure equations (see \citealt{van92}), which
is designed to follow the evolution of a very thin H-burning shell along the RGB
very efficiently.  Indeed, the entire track from the base of the giant branch
until the onset of the helium flash, which is the only part of the evolution of
a star that utilizes this technique, can be computed in less than 0.5\% of the
{\it cpu} time required by codes that take mass to be the independent variable.]

\begin{figure}[t]
\plotone{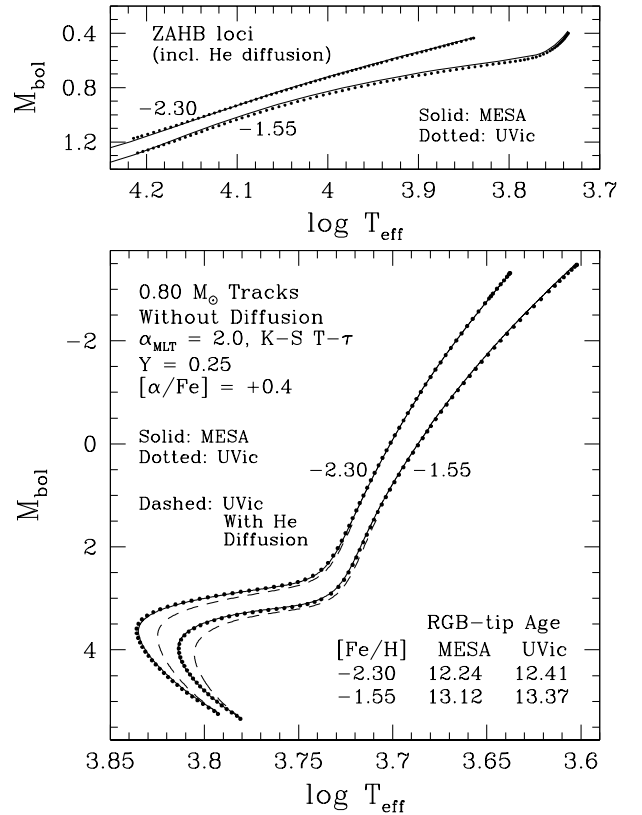}
\caption{{\it Bottom panel}: comparison of MESA and Victoria tracks for the MS
and RGB phases of stars having the indicated mass and chemical compositions.  As
noted adjacent to the giant branches, the adopted [Fe/H] values are $-2.30$ and
$-1.55$.  Because the two codes employ different treatments of diffusion, and
since only the Victoria code allows for extra mixing processes below convective
envelopes to limit the efficiency of gravitational settling, we opted to
intercompare non-diffusive tracks.  However, the effects of diffusion and extra
mixing are included in the dashed loci.  (For the sake of clarity, these 
tracks have been plotted only for $M_{\rm bol} > 1.5$.  All models assume a
value of $\amlt = 2.0$ for the usual convective mixing-length parameter, and
the surface pressure boundary conditions were obtained by integrating the
hydrostatic equation together with the atmospheric $T$--$\tau$ relation given by
\citet{ks66}.  {\it Top panel}: comparison of fully consistent ZAHB loci.
Diffusion was treated in the precursor models because the inclusion, or not, of
this physics has important consequences for the predicted abundances of helium
in the envelopes of HB stars, and therefore for the luminosities of the latter
and consequent ZAHB-based distance moduli.  The MESA and Victoria codes predict
nearly the same helium core masses and envelope abundances in RGB tip stars,
which explains why the respective ZAHBs agree so well.}
\end{figure}
 
It turns out that, as illustrated in Figure 1, there is excellent consistency
between MESA and Victoria tracks and ZAHB loci.  Both sets of calculations
assumed exactly the same abundances of helium ($Y = 0.25$) and the heavier
elements; specifically, the solar metals mixture given by \citet{ags09}, with a
0.4 dex enhancement of the $\alpha$-elements, then scaled to [Fe/H] $= -1.55$
and $-2.30$ (as indicated).  Since this mix of the heavy elements had been
previously considered by V14, we were able to use the same opacities
that had been generated for that project via the Livermore Laboratory OPAL
opacity Web site\footnote{http://opalopacity.llnl.gov} and those calculated
using the code described by \citet{faa05} for high- and low-temperatures,
respectively.  In addition, {\it for this particular comparison}, the preferred
rates from the JINA {\it Reaclib} database (\citealt{caf10}) for the most
important H- and He-burning nuclear reactions were incorporated into the
Victoria code so that this component of the stellar physics would be identical
to the treatment adopted in the very recent version of the MESA code
(specifically, release 7624) that has been used throughout this investigation.  

Although MESA has a large number of parameters that provide the means to control
the speed and accuracy of the model computations, and to choose among different
prescriptions for the equation of state, the nuclear reaction network, the 
reaction rates, etc., we used default values of all, but one, of these
parameters.  The best agreement with Victoria stellar models is obtained if
cubic interpolations of the opacities with respect to $Z$ are adopted instead
of quadratic interpolations (the default option).  (In the Victoria code, cubic
splines are employed to evaluate the opacities at different values of $Z$.)  For
consistency, we chose the ``Krishna-Swamy" option (see \citealt{ks66}) for the
atmospheric $T$--$\tau$\ relation, as well as the ``Henyey" option
(\citealt{hvb65}) for the treatment of the mixing-length theory of convection,
with the mixing-length equal to 2.0 pressure scale-heights.  This is very close
to the value found from a Standard Solar Model (see V14). 

In comparison with the models computed by V14, the ``Victoria" tracks 
that appear in Fig.~1 are cooler by only $\delta\log\teff \approx 0.0008$,
while predicting the same RGB-tip age to within 0.02 Gyr.  The adoption in the
published 2014 models of a slightly reduced rate (from \citealt{mfg08}) for the
$^{14}$N$(p,\,\gamma)^{15}$O reaction, as compared with the JINA rate for this
reaction, also has minor consequences for ZAHB models, in that the helium core
mass at the top of the giant branch is reduced by $\sim 0.007\,\msol$, which
translates to a lower luminosity by $\sim 0.015$ mag at a fixed $\teff$\ on
the HB (when differences in the model $\teff$ scale are also taken into
account).  Thus, for instance, the ZAHB-based distance moduli derived by V13
would have been reduced by $\approx 0.015$ mag, implying increased ages by $\sim
0.15$ Gyr, had their models been based on the JINA nuclear reaction rates
(\citealt{caf10}) instead of the adopted ones.  Be that as it may, Fig.~1 shows
that the evolutionary tracks and ZAHBs computed by the MESA and Victoria codes
are in excellent agreement when both employ very close to the same physics.
This figure provides ample justification for combining MESA models for the HB
phase with Victoria isochrones for the MS and RGB phases.  

The prediction of slightly higher ages by the Victoria code (by $\lta 2$\%,
see Fig.~1) appears to be due mostly to small differences in the respective
equation-of-state (EOS) formulations, though differences in, e.g., some of the
numerical methods that are used could be part of the explanation.  Exploratory
computations that we carried out revealed that most of this difference would be
eliminated if we used the EOS developed by A.~Irwin, widely known as
``FreeEOS"\footnote{http://freeeos.sourceforge.net}, to generate the Victoria
track instead of the default EOS (see \citealt{vsr00}).  The latter is normally
favored because it is computationally much faster than FreeEOS (by at least a
factor of 3 if the EOS4 implementation of FreeEOS is employed, and by much
larger factors if EOS1--EOS3 are used).  This is an
important advantage when generating large grids of tracks and isochrones.
Errors at the level of $\sim 2$\% are, anyway, much smaller than those
associated with current distance and metal abundance (especially [O/Fe])
determinations.

It is worth mentioning that MESA can follow the evolution of a track through
the core Helium Flash all the way to the HB (and beyond), which requires several
thousand stellar models.  Indeed, the most massive ZAHB model is always created
in this way.  Mass is then removed from the envelope of this initial model, in
small increments, to generate lower mass ZAHB models.  The Victoria code, on
the other hand, inserts into a previously converged ZAHB structure the chemical
abundance profiles from an appropriate red-giant precursor (one in which the
He-burning luminosity has exceeded $100 L_\odot$), and then relaxes that
structure via many short timesteps until the central He abundance has decreased
by $\delta Y \approx 0.05$ from an initial value of $1-Z$, where $Z$ is the
total mass-fraction abundance of the metals.  This endpoint is suggested by
MESA models that have been evolved through the Helium Flash.  It is
just a matter of repeating this procedure, on the assumption of the same helium
core mass but different envelope masses, until an entire ZAHB extending to,
say, $\log\teff = 4.20$ has been generated.  As shown in the upper panel of
Fig.~1, this classical, computationally much less demanding approach (see, e.g.,
\citealt{dor92}) works extremely well if executed carefully.  (For a discussion
of the methods that have been used to compute ZAHB models, see \citealt{sw05}.)

The subsequent evolution of low-mass HB stars is known to be strongly dependent
on the treatment of mixing at the boundary of the convective helium core (e.g.,
\citealt[and references therein]{sdi03}).  Because C-rich material below that
boundary has a higher opacity than the He-rich matter above it, a discontinuity
is created in the ratio of the radiative and adiabatic temperature gradients,
$\nabla_{\rm rad}/\nabla_{\rm ad}$, at the boundary.  As the core grows
in mass and becomes more enriched in carbon, this ratio can exceed 1.0 at the
boundary, while the minimum value inside the core falls below unity.  Such a
variation of $\nabla_{\rm rad}/\nabla_{\rm ad}$ with mass implies that this
region will split into a smaller convective core and a surrounding zone that
undergoes semi-convective mixing.  Unfortunately, precisely how this mixing
occurs is still an open question due to the lack of suitable 3D hydrodynamical
simulations that treat all of the relevant microphysics (e.g., nuclear
reactions, opacity variations) on a thermal timescale.
 
In the absence of such simulations, a number of different mixing prescriptions,
considered to be reasonable, have been developed for use in post-ZAHB models in
the hope that reasonable consistency with observational constraints would be
found.  Let it suffice it to say that \citet{ccc15} have recently concluded
that their proposed ``maximal overshoot" treatment of mixing in convective cores
results in stellar structures whose non-radial pulsations appear to match those
of field HB stars, as derived from {\it Kepler} observations, better than those
computed for models that have implemented other mixing prescriptions.  Based on
these findings, we have fine-tuned the values of the parameters that control
convective overshooting in the MESA code so that our models for the HB phase
have evolving He abundance profiles that closely resemble those reported by
\citet{ccc15} for their ``maximal overshoot" case.  (A full accounting of what
we have done, supported by relevant plots, will be provided in a later paper in
this series by P.~Denissenkov et al.  The same paper will make the grids of HB
tracks used in this investigation available to the astronomical community.)
Compared with models that neglect core overshooting, our models predict more
massive He cores and longer core He-burning lifetimes ($\sim 100$ Myr) by
nearly a factor of two.  In addition, our evolutionary tracks do not contain
loops caused by so-called ``core breathing pulses", in good agreement with the
most recent estimates of the R$_2$ parameter that measures the relative
lifetimes of asymptotic-giant-branch and HB stars (see \citealt{ccl16}).  

\section{Analysis of GC Observations}
\label{sec:obs}

Since the main goal of this investigation is to obtain (if possible) fully
consistent interpretations of the MSTO and HB populations in M\,3, M\,15, and
M\,92, our analysis of each cluster begins by determining its
distance and age.  To accomplish this, all of the observed colors are first
dereddened, assuming an estimate of $E(B-V)$ that is supported by analyses of
dust maps (\citealt{sfd98}, \citealt{sf11}).  For colors other than
$B-V$, we have used $E(\zeta - \eta) = (R_\zeta - R_\eta)E(B-V)$, where
$R_\zeta$ and $R_\eta$ have the values given by CV14 (see their Table A1) for  
filters $\zeta$\ and $\eta$.  Then, to determine the apparent distance
modulus, the observed magnitudes are adjusted until the lower bound of the
distribution of member HB stars coincides with a ZAHB that has been computed
for an adopted value of $Y$, and for assumed metal abundances that are 
consistent with recent spectroscopic results.  Having set the value of
$(m-M)_V$ in this way, it is a straightforward matter to fit isochrones for the
same initial chemical abundances as the ZAHB to the turnoff photometry in order
to derive the corresponding age.  (It has already been shown by V13 that
current ZAHB loci reproduce the morphologies of observed HBs very well,
especially in the case of GCs that have [Fe/H] $\lta -1.0$, and that they
seem to be very good distance indicators.)

To complete our analysis, the full grid of HB evolutionary tracks on which the
ZAHB locus was based is overlaid onto the observed HB population.  Via suitable
interpolations within this grid, the effective temperatures, luminosities, and
masses that correspond to published determinations of the mean magnitudes and
colors of the RR Lyrae variables (i.e., the properties of equivalent ``static
stars") are determined.  This information, together with the value of $Z$ that
was assumed in the model computations, enable one to calculate the periods,
in units of days, of the $ab$-type (fundamental mode) and $c$-type
(first overtone) pulsators using the equations (from \citealt{mcb15}):
\begin{align}
\log\,P_{ab} &= (11.347 \pm 0.006) + (0.860 \pm 0.003)\log(L/L_\odot)\nonumber\\ 
 &- (0.58 \pm 0.02)\log({\cal{M}/\cal{M}_\odot})\\
 &- (3.43 \pm 0.01)\log\teff + (0.024 \pm 0.002)\log\,Z\nonumber
\end{align} 
\noindent and
\begin{align}
\log\,P_c &= (11.167 \pm 0.002) + (0.822 \pm 0.004)\log(L/L_\odot)\nonumber\\
 &- (0.56 \pm 0.02)\log({\cal{M}/\cal{M}_\odot})\\
 &- (3.40 \pm 0.03)\log\teff + (0.013 \pm 0.002)\log\,Z.\nonumber
\end{align}
(These results were derived from state-of-the-art hydrodynamical models of
RR Lyrae variables that employ a nonlinear, nonlocal, time-dependent treatment
of convection.)  Once the periods predicted by the stellar models have been
determined, they are compared with the observed periods on a star-by-star basis.

It can be anticipated from the preceding remarks that several plots have been
prepared for each cluster, and indeed, we now turn to a presentation and
discussion of these plots.  We begin with M\,3, mainly because an analysis of
its CMD and RR Lyrae population appears to be relatively free of difficulties,
and end with M\,15, which poses a much greater challenge than either M\,3 or
M\,92.

\subsection{M\,3}
\label{subsec:m3}

\begin{figure}[t]
\plotone{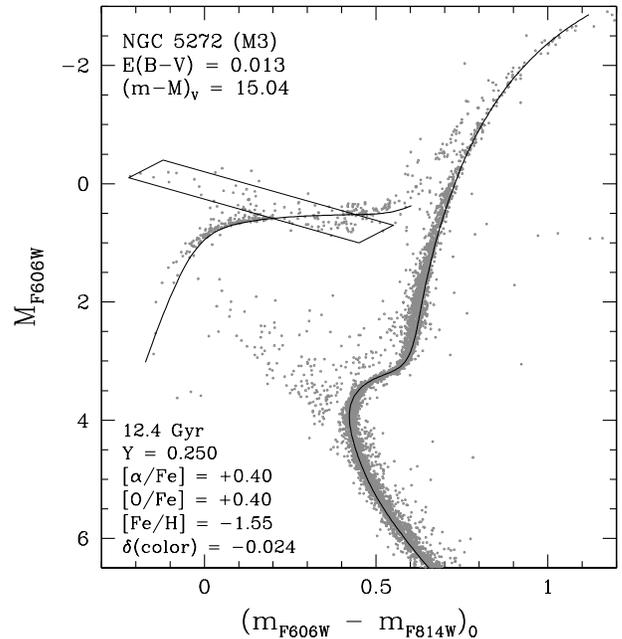}
\caption{Fit of a 12.4 Gyr isochrone for the indicated chemical abundances to
the turnoff and subgiant photometry of M\,3, assuming $E(B-V) = 0.013$
and $(m-M)_V = 15.04$, as found from the ZAHB models.  The apparent distance
modulus in the $F606W$ magnitude was calculated from the relation
$(m-M)_{F606W} = (m-M)_V - 0.246E(B-V)$; see CV14.  (Note that RR Lyrae 
at different phases of their pulsation cycles are responsible for most
of the scatter of points in the region contained within the parallelogram.
Because the {\it HST} observations were taken as part of a snap-shot survey,
magnitude- or intensity-weighted mean magnitudes cannot be calculated for the
variable stars from those data.)  To reproduce the TO color, the isochrone had
to be adjusted by 0.024 mag to the blue.}
\end{figure}

As it is usually worthwhile to examine fits of isochrones to as many different
CMDs as possible, we have opted to consider both the {\it HST} photometry of 
M\,3 that was obtained by \citet{sbc07} and the latest calibration of
ground-based $BVI_C$ data by P.~Stetson (as described, and used, in the study
by \citealt{vsb15}).  A plot of the $F606W, F814W$ observations is
shown in Figure 2, which illustrates that a ZAHB for the indicated chemical
abundances provides quite a good match to the lower bound of the distribution
of non-variable HB stars at $(m_{F606W} - m_{F814W})_0 \lta 0.2$ and $\gta 0.4$.
The isochrone, for the same abundances, that provides the best fit to the TO
and SGB is one for an age of $\approx 12.4$ Gyr.  While a small color offset had
to be applied to the isochrone in order to match the observed turnoff
color, this has no impact on the inferred age (see V13).  It does indicate,
however, that there must be a small problem with, e.g., the model $\teff$
scale, the adopted color transformations, the photometric zero-points, and/or
the assumed chemical composition.  Regardless, the level of agreement between
theory and observations is quite satisfactory when the adopted or derived
properties of M\,3 are close to currently favored values (see, e.g., the 
entries for this GC in the latest edition of the catalog by
\citealt[see our footnote 4]{har96}). 

The same can be said of Figure 3, which is identical to Fig.~2 except that the
isochrones are compared with the principal photometric sequences of M\,3 on
the $[(V-I_C)_0,\,M_V]$-plane.  Interestingly, the predicted and observed
turnoff colors agree to within 0.002 mag, but the cluster RGB is offset to the
blue by a larger amount than in the previous plot.  Because of the many factors
that play a role in such comparisons, it is not easy to determine which one is
mostly responsible for these discrepancies.  It seems unlikely that they can
be attributed primarily to errors in the predicted temperatures because any
$\teff$\ adjustments that eliminate the problems in one CMD will exacerbate the
difficulties in the other CMD --- especially in view of the similarity between
$(F606W,\,F814W)$\ and Johnson-Cousins $(V,\,I_C)$.  Aside from small zero-point errors, the
photometry is probably quite reliable in a systematic sense, but this may not 
be true of current color--$\teff$ relations.  In any case, it is very
encouraging to find that the quality of the fits to both the HB and the TO
observations are comparable in Figs.~2 and 3.

\begin{figure}[t]
\plotone{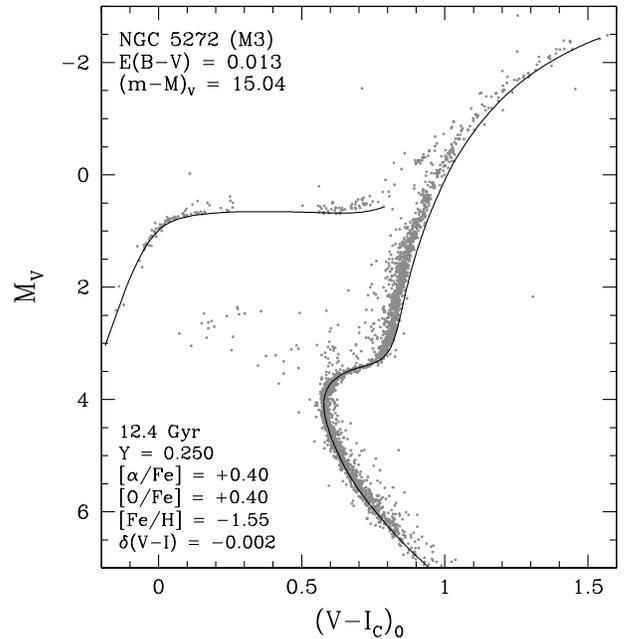}
\caption{Similar to the previous figure, except that the stellar models are
compared with $VI_C$ observations of M\,3.}
\end{figure}

Remarkably, of the three CMDs that have been considered, the same stellar
models provide the best match to the $[(B-V)_0,\,M_V]$-diagram of M\,3, as
shown in Figure 4.  This is unexpected because the blanketing is more severe,
and hence more problematic from the modeling perspective, in the $B$ bandpass
than at longer wavelengths.  Figs.~2--4 thus demonstrate that inconsistencies
in predicted colors at the level of a few hundredths of a magnitude, especially
for cool stars, are unavoidable.  However, the ZAHB-based {\it apparent}
distance moduli and predicted ages are largely independent of color-related
uncertainties.  It is worth mentioning that \citet{vsb15} used the same
isochrones, but different ZAHB models, in their fits to the same $BV$ photometry
of M\,3.  They obtained an age of 12.25 Gyr, which is slightly younger than our
determination (12.4 Gyr), because they adopted a slightly larger value of
$(m-M)_V$.  An even younger age (11.75 Gyr) was derived by V13 in their survey
of GC ages, due mainly to their use of stellar models that assumed a
significantly higher abundance of oxygen, which more than compensated for a
reduced distance modulus.

\begin{figure}[t]
\plotone{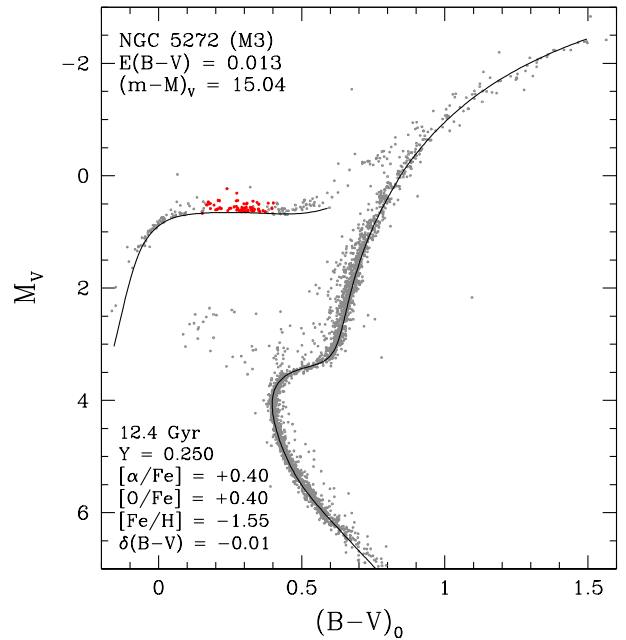}
\caption{Similar to the previous two figures, except that the stellar models are
compared with $BV$ observations of M\,3.  The ``static equivalent" properties 
of the sample of RR Lyrae stars that are considered in this paper (see the text)
have been plotted as small red filled circles.}
\end{figure}

The RR Lyrae that appear in Fig.~4 as red dots were taken from the study by
\citet[their Tables 1 and 2]{ccc05}.  All variables that were flagged as having
large scatter in their light curves or low amplitudes (a possible sign of
blends), or which exhibited some evidence for the presence of companions or for
the Blazhko effect (see, e.g., \citealt[and references therein]{bk11}), were
removed from the sample.  However, even when such strict selection criteria are
adopted --- which we can afford to employ in the case of such an RR Lyrae-rich
cluster as M\,3 --- we are still left with a total sample of 69 variables, 46
of which are fundamental ($ab$-type) pulsators and 23 of which are
first-overtone ($c$-type) pulsators. 

\citet{ccc05} converted colors (but not magnitudes) to
their static equivalents, based on the prescriptions given by \citet{bcs95}.
(Fortunately, the differences between the static colors so derived and
magnitude-weighted mean colors are typically $\lta 0.02$ mag.)  By interpolating
in the Bono et al.~tables, we were able to compute the static $V$ magnitudes
for the M\,3 RR Lyrae.  It turns out that they generally agree to within
$\sim 0.002$ mag with the mean magnitudes given by Cacciari et al., who
integrated the light curves in intensity and then converted the resultant
integrations to magnitudes.  Accordingly, we have simply adopted the values of
$\langle V\rangle_i$ and $(B-V)_S$ that are tabulated by Cacciari et al.

\begin{figure}[t]
\plotone{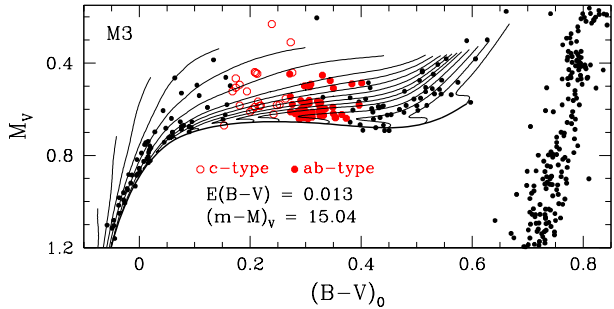}
\caption{Overlay of evolutionary tracks for core He-burning stars and the same
ZAHB that appears in the previous figure onto the CMD for the HB and RGB
populations of M\,3 that have $0.1 \le M_V \le 1.4$.  Filled and open circles
(in red) identify the $ab$-type and $c$-type RR Lyrae, respectively.}
\end{figure}

Figure 5 focuses in on the region of the CMD that contains the RR Lyrae and
non-variable HB stars of M\,3, as well as cluster giants that lie within the
same range of $M_V$.  The stars and ZAHB that appeared
in the previous figure are reproduced here, but different symbols are used to
identify the fundamental and first-overtone pulsators (as noted).  A grid of
post-ZAHB tracks, for the same initial chemical abundances that were assumed in
the isochrones (see Figs.~2--4) and for masses in the range of 0.80--0.58
$\msol$\ (in the direction from red to blue colors) has been superimposed on
the observations.  They begin at the ZAHB and end when the central helium
abundance has fallen to $Y_C \sim 0.01$, which typically takes $\sim 90$\ Myr.

Except for the four most massive HB models, evolutionary sequences were computed
for masses that differed by $0.005 \msol$\ in the vicinity of the instability
strip, rising to $0.01 \msol$\ for the hottest models.  This spacing is 
sufficiently fine that precise predictions of the masses, luminosities, and
effective temperatures of the RR Lyrae stars can be obtained simply by linear
interpolations within the grid (or by extrapolating just outside of it, in the
case of the brightest variables).  Since the stellar models were computed for
$Z = 7.623\times 10^{-4}$, the periods of the variables can be calculated 
using equations (1) and (2), and then compared with the observed periods. 

The results of this exercise are better than one might have expected (as we will
show shortly), though the computed periods for the $ab$-type variables
tend to be somewhat too low.  The most likely explanations of this problem are
(i) the predicted temperatures are too high --- despite the
fact that isochrones need to be shifted to the blue to match the turnoff color,
which goes in the opposite direction, (ii) the values of $(B-V)_S$ given by
\citet{ccc05} are too blue, or (iii) the coefficient that multiplies
$\log\teff$ should be reduced (in an absolute sense).  It is well known that
the temperatures of stellar models are much more uncertain than their
luminosities, and $\teff$\ uncertainties will obviously have a much bigger
impact on the calculated periods of RR Lyrae than those associated with
luminosities or masses.  

\begin{figure}[t]
\plotone{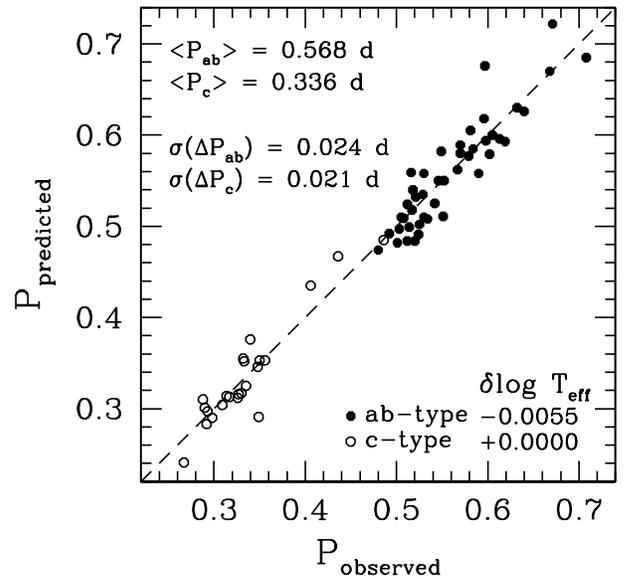}
\caption{Comparison of the observed periods, in days, of the $ab$- and $c$-type
RR Lyrae in M\,3 with those derived from the evolutionary tracks shown in the
previous figure (for $Y = 0.25$, [Fe/H] $= -1.55$, and [$\alpha$/Fe] $= +0.40$).
The observed periods have the mean values that are given in the top,
left-hand corner of the plot.  The same values of \pab\ and \pc\ are obtained
for the predicted periods if the temperatures of the variables are adjusted by
the amounts specified in the lower right-hand corner (see the text).
\citet{ccc05} estimate the internal errors in their $(B-V)_S$ colors to be
typically 0.02 mag, which translates to errors in $\log\teff$ and the predicted
periods, in turn, of $\approx 0.007$ and $\sim 0.03$~d.  The measured periods
are known to better than $\pm 0.00001$~d.  The differences between the predicted
and the observed periods have a standard deviation $\sigma = 0.024$~d and
0.021~d in the case of the $ab$- and $c$-type variables, respectively.}
\end{figure}

In fact, rather good consistency between the
predicted and observed periods, and the corresponding value of
\pab,\footnote{We are using \pab\ and \pc\ to represent the average periods,
either predicted or observed, of the selected samples of cluster RR Lyrae stars.
To properly predict the mean periods, one should compute synthetic HBs --- in
which case, consistency between theory and observations would depend on how
well the tracks are able to explain the observed {\it distributions} of the
variables, in addition to reproducing their masses, luminosities, and
temperatures.  Simulations that take evolutionary speeds and the predicted
locations of the boundaries of the instability strip into account will be
presented in Paper II.} can be obtained if $-3.425$ is adopted instead 
of $-3.430$ for the $\log\teff$ coefficient, which has a $1\sigma$ uncertainty
of $\pm 0.01$ according to equation (1).  However, the periods given by
period--mean-density relations involve relatively small uncertainties.  That is,
changes to the various coefficients and the zero point in different versions of
such equations tend to compensate for one another so as to yield nearly the
same periods; for some discussion of this point, see \citet{cat93}.  As a
result, it is unlikely that the $\teff$\ coefficients can be altered in
equations (1) and (2) without concomitant changes to other coefficients.

For this reason, it is preferable to correct the predicted $\teff$\ scale when
attempting to match the observed values of \pab\ and \pc. (Doing so serves
to compensate for errors in the adopted values of $(B-V)_S$, the color--$\teff$
relations that are used, and the temperatures of the stellar models.)  In
Figure 6, the observed periods of the selected M\,3 RR Lyrae stars are compared
with those computed using equations (1) and (2) after the temperatures derived
for them via interpolations in the grid of HB tracks shown in Fig.~5 have
been adjusted by the amounts specified in the lower right-hand corner.  With
these adjustments, the calculated values of \pab\ and \pc\ reproduce the
observed values (given in the upper left-hand corner of the plot) to three
decimal places.  This consistency was achieved simply by iterating on the
relevant $\delta\log\teff$\ values.  Note that a temperature reduction that
was applied to the fundamental-mode pulsators has the effect of increasing the
calculated period of an RR Lyrae that has $P_{ab} = 0.55$~d~by $\approx
0.024$~d. (Changes to the temperatures, luminosities, or masses that are
predicted for a given RR Lyrae will move the point representing that star
vertically up or down in Fig.~6 at the observed value of $\log P$.  For
instance, the two $c$-type variables that lie above the dashed line with
observed periods of $\sim 0.42$~d~would shift onto that line if their values of
$\log\teff$, $M_{\rm bol}$, or mass were increased by 0.009 dex, 0.093 mag, or
$0.08\,\msol$, respectively.)   

\begin{figure}[t]
\plotone{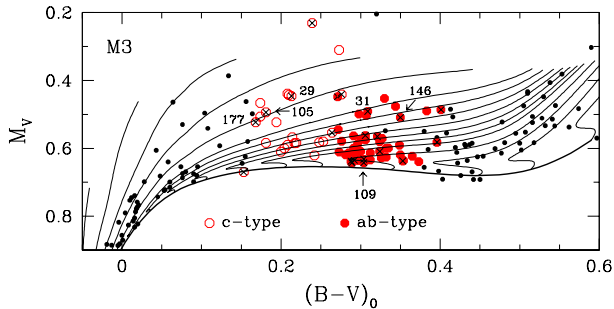}
\caption{A somewhat magnified version of Fig.~5 in which the RR Lyrae with
$\vert\Delta P_{ab}\vert > 0.024$~d and $\vert\Delta P_c\vert > 0.021$~~d
are identified by crosses superimposed on the relevant open or filled circles.
Variable identification numbers are specified only for those stars that are
referenced explicitly in the text.}
\end{figure}

The {\it dispersion} in the predicted periods relative to the observed
periods is presumably due mostly to errors in the values of $(B-V)_S$ that were
determined by \citet{ccc05}, given that the HB evolutionary tracks are expected
to be quite robust in a differential sense.  Support for this assertion is
provided in Figure 7, which shows a somewhat magnified version of Fig.~5 in
which the stars with $\vert\Delta P_{ab}\vert > 0.024$~d and
$\vert\Delta P_c\vert > 0.021$~d (i.e., the points furthest from the ``line
of equality" in Fig.~6) are identified by crosses.  The majority of them are
located in close proximity to stars for which the predicted and observed periods
are in good agreement.  This is certainly true of V109 and other crossed
variables in the dense concentration of $ab$-type RR Lyrae at $M_V \sim 0.6$
and $(B-V)_0 \sim 0.3$, but the same thing is found elsewhere in Fig.~7.  For
instance, the calculated periods of V105 and V177 differ, in turn, by
$+0.022$~d and $-0.058$~d from their measured values, though the difference is
only $+0.011$~d for the variable that is located between V105 and V177.
Similarly, V29 and V31 have nearly the same CMD locations as other variables in
which the predicted and observed periods agree to within $\pm 0.012$~d.  In
fact, consistency at this level is obtained for approximately half of the RR
Lyrae stars in our sample.  [Were we to drop from consideration the most
discrepant points (i.e., the stars denoted by crosses in Fig.~7), the
differences between the calculated and measured periods for the resultant sample
of 32 $ab$-type and 16 $c$-type variables would have dispersions with
$\sigma(\Delta P_{ab}) = 0.013$~d and $\sigma(\Delta P_c) = 0.009$~d.]

This is really very encouraging consistency between theory and observations
given that such differences correspond to errors of $\lta \pm 0.004$ in
the values of $\log\teff$ that are derived for the variables from the HB tracks
and the adopted color--$\teff$\ relations (by CV14).  The fact that the
most problematic stars are roughly evenly distributed as functions of both
magnitude and color, especially in the case of the fundamental-mode variables,
suggests that the $\Delta P$ dispersions are primarily statistical fluctuations
rather than, say, the consequence of chemical abundance variations (though the
latter could be contributing factors).  Note that the star with the largest
difference between the predicted and observed period (0.080~d) is V146, which
lies close to the middle of the color range spanned by the $ab$-type RR Lyrae. 

Predicted luminosities also appear to be quite reliable.  If $(m-M)_V = 15.04$
is assumed for M\,3 (see Figs.~2--4), \vavg\ $= 0.583$ is obtained for the
entire sample of $ab$-type variables that we have considered.  According to
\citet{cgb03}, the fundamental-mode pulsators residing in the Large Magellanic
Cloud (LMC) have $\langle V_0\rangle = 0.214($[Fe/H]$ + 1.5) + 19.064$.  On the
assumption of the accurate eclipsing-binary distance derived by \citet{pgg13}
for the LMC, which corresponds to $(m-M)_0 = 18.494$, the Clementini et
al.~relation yields \vavg $= 0.559$ for RR Lyrae that have [Fe/H] $= -1.55$
(the metallicity that we have adopted for M\,3).  This differs from our
determination by only 0.024 mag, which is well within distance modulus and
metallicity uncertainties (both for M\,3 and the LMC variables).  On the other
hand, we could easily obtain a brighter value of \vavg\ simply by adopting a
slightly higher helium abundance or a lower [Fe/H] value.  Indeed, a metallicity
close to $-1.7$ (recall the work of \citealt{zw84}), or less, is well within the
realm of possibility, especially as there has been some movement in recent
spectroscopic investigations towards lower metallicities for metal-poor GCs
(e.g., see \citealt{sks11}, \citealt{rs11}, \citealt{rt15}).

Although a reinvestigation of M\,3 using the same methods and codes that were
employed in the aforementioned studies has yet to be carried out, a lower [Fe/H]
value would help to alleviate a possible problem with the interpretation of the
M\,3 HB shown in Figs.~5 and 7 by reducing the extent of the post-ZAHB blue
loops.  As discussed by \citet[see his Figs.~8, 9]{san81}, one would expect to
see some overlap of the colors of $ab$- and $c$-type variables, as a result of
the hysteresis effect (\citealt{vb73}), if tracks with blue loops accurately
describe the evolution of the observed HB stars.  This is not a major problem
for the models plotted in Figs.~5 and 7 because the lengths of the blue loops
amount to no more than $\delta (B-V)\sim 0.05$ mag at the color which separates
$c$- and $ab$-type variables, but the observations indicate that there is
very little, if any, overlap whatsoever of the fundamental and first-overtone
pulsators --- at least for the sample of RR Lyrae that we have considered.

The limited work that we have done on this issue so far indicates that the
lengths of blue loops, in the vicinity of the instability strip, decrease
relatively slowly with [Fe/H]; i.e., they would still be present, but shortened
by, e.g., $\sim 25$\% at [Fe/H] $= -1.7$ (assuming constant $Y$ and
[$\alpha$/Fe]).  Higher helium abundances would exacerbate this problem (see
below), but a small reduction in $Y$ and/or [CNO/Fe] (or an increased helium
core mass; see \citealt{cat92}) would have beneficial
consequences in this regard.  It is worth mentioning that a modest decrease in
the assumed [Fe/H] value produces no more than minor changes to the effective
temperatures and masses that are derived from the corresponding HB models.  
Thus, we would have obtained a plot that is very similar to Fig.~6 had we
adopted a lower [Fe/H] value for M\,3 by, e.g., $\sim 0.15$ dex (while 
retaining the same values of the other chemical abundance parameters).

\begin{figure}[t]
\plotone{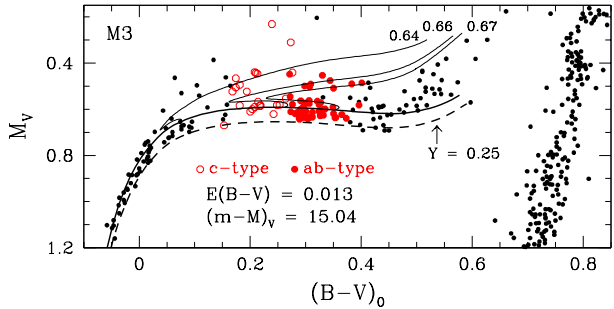}
\caption{Similar to Fig.~5, except that a ZAHB for $Y = 0.27$ and associated
evolutionary tracks for 0.64, 0.66, and 0.67 $\msol$ have been superimposed
onto the observed HB stars in M\,3.  The location of the same ZAHB (for $Y =
0.25$) that appears in Fig.~5 is reproduced here as the dashed curve.  Note that
exactly the same reddening and distance modulus have been assumed in both
figures.}
\end{figure}

The close matches of a ZAHB for constant $Y$ to the faintest HB stars over the
entire ranges in color plotted in Figs.~2--5 provide a strong argument that at
least the lowest luminosity HB stars in M\,3 have nearly the same helium
abundance.  (The same conclusion was reached, based on similar findings, by
\citealt[also see \citealt{vca16}]{cgs09}.)  That our computations preclude
variations of $Y$ by more than $\sim 0.005$ in {\it these} stars is demonstrated
in Figure~8, which shows that the displacement {\it at any color} between the
faintest HB stars and the ZAHB for $Y = 0.25$ is a small fraction of the
separation between ZAHBs for $Y = 0.25$ and 0.27.  

These results argue against
the explanation of the Oosterhoff dichotomy recently
proposed by \citet{jlj14}.  In their scenario, the RR Lyrae in M\,3 are
expected to have lower helium abundances than the non-variable stars on either
the red or blue sides of the instability strip, which should cause the latter to
be somewhat brighter than the ZAHB that is relevant for the variable stars.
(If anything, the faintest RR Lyrae appear to be slightly brighter than the
non-variable stars to the left or right of them, but this could be the result of
small zero-point differences in the photometry for the variable and non-variable
stars, which come from different sources.)

However, Fig.~8 does not rule out the possibility that some fraction of the
stars lying above the $Y = 0.25$ ZAHB have higher helium abundances, including
some of the brightest $c$-type variables, judging from their locations relative
to the track for $Y = 0.27$ and ${\cal M} = 0.64\,\msol$.  [Unfortunately, it is
not possible to use the predicted periods to constrain the helium abundances of
the RR Lyrae because the only quantity that varies appreciably with $Y$ at a
given CMD location, {\it assuming fixed values of the reddening and distance
modulus}, is the mass, and its variation ($\sim 0.01$--0.03 $\msol$\ for
$\delta Y \lta 0.02$) has only a small effect on the period; see equations (1)
and (2).] 

As mentioned above, the apparent lack of any overlap of the colors of the
$ab$- and $c$-type variables implies that stars which began their core
He-burning lifetimes as fundamental-mode pulsators do not follow tracks that
have blue loops or the blue loops are too small to reach very far into the
region of the instability strip where only first-overtone pulsators are found
(see Fig.~8 by \citealt{san81}).  Alternatively, the hysteresis mechanism
does not occur in real stars.  Since these loops are obviously quite
a strong function of $Y$ (compare Figs.~7 and 8), a helium abundance slightly
less than $Y = 0.25$ (but within the uncertainties of the primordial helium
abundance; see \citealt{ksd11}) and/or some refinement of the assumed CNO
content would appear to be necessary to explain the sharp boundary between the
fundamental and first-overtone pulsators at $(B-V)_0 \approx 0.27$.  (Some
additional discussion of this point is given in \S~\ref{sec:sum}.)  In any
case, our analysis suggests that most of the stars in M\,3 have nearly the same
helium abundance, though star-to-star variations as large as $\delta Y \sim
0.02$ cannot be ruled out.

As already mentioned, further constraints on the properties of M\,3 may be
obtained from a consideration of synthetic HB populations, but we defer such
work to the next paper in the current series, which will be devoted to a study
of M\,3 and M\,13.

\subsection{M\,92}
\label{subsec:m92}

Although most investigations over the years have found that M\,92 has [Fe/H]
$\sim -2.3$ (e.g., \citealt{zw84}, \citealt{spk00}, \citealt{be03},
\citealt{cbg09a}), lower values by 0.2--0.4 dex have been obtained in some
spectroscopic studies (e.g., \citealt{pkc90}, \citealt{ksb98}), including the
recent one by \citet{rs11}.  In view of this, we decided to fit stellar models
for [Fe/H] $= -2.30$ and $-2.60$ to the CMD of M\,92, and to the properties of
its variable stars, in order to determine whether they indicate any preference
for one of these metallicities over the other.  

The best available photometry for the cluster RR Lyrae is given by
\citet{kop01}, who derived intensity-weighted mean $\langle V\rangle_i$
brightnesses and magnitude-weighted $(V-I_C)_m$ color indices, as calculated
from the difference in the magnitude-weighted magnitudes $(V)_m$ and $(I_C)_m$,
for the variables.  We have therefore used $[(V-I_C)_0,\,M_V]$-diagrams 
throughout our study of M\,92.  However, we did verify that the ZAHB and
best-fit isochrone on this CMD provide equally good interpretations of {\it HST}
$F606W,\,F814W$ and $B,\,V$ data for the TO and HB stars.  [These plots have not
been included here because they merely serve to confirm what has already been
demonstrated in Figs.~2--4 for M\,3; namely, that small CMD-dependent zero-point
and systematic offsets between predicted and observed colors are commonly found
--- though they do not affect the derived distance modulus and age.]
  
M\,92 is known to have 17 RR Lyrae (\citealt{kop01}), but only 12 of them
(8 $ab$-types and 4 $c$-types) have reliable measured magnitudes according to
the online version of the \citet{cmd01} catalog of variable stars in
GCs\footnote{http://www.astro.utoronto.ca/$\sim$cclement/read.html}.  The
properties of one of the remaining fundamental pulsators (specifically, V6) seem
suspect as well because it has a relatively short period (0.600 d) despite being
the most luminous RR Lyrae ($\langle V\rangle_i = 14.96$) and having a color
(and therefore $\teff$) that is very similar to those of the other $ab$-type
variables.  By comparison, V1 has $\langle V\rangle_i = 15.02$ and a period of
0.703 d.  Because an unreasonably large mass would have to be invoked in order
to explain the period of V6 using equation (1) if the values of
$\langle V\rangle_i$ and $(V-I_C)_m$ given by Kopacki for this
star are adopted, something is clearly awry.  For this reason, V6 has been
dropped from further consideration.

\subsubsection{Isochrones, ZAHBs, and RR Lyrae}
\label{subsubsec:modm92}

The bottom panel of Figure 9 shows that a ZAHB for [Fe/H] $= -2.30$,
[$\alpha$/Fe] $= 0.4$ (e.g., \citealt{car96}, \citealt{spk00}), and $Y = 0.250$
provides quite a good fit to M\,92's faintest, non-variable blue HB stars if
$(m-M)_V = 14.74$ and the foreground reddening is $E(B-V) = 0.023$ mag.  This
value of $Y$ is within the uncertainties associated with current estimates of  
the primodial abundance of helium and the abundances that have been
derived from helium lines in the spectra of HB stars in M\,30 and NGC\,6397
with $\teff\sim 10^4$~K (see \citealt[as well as references therein]{mll14}).
(M\,92 and M\,30 probably have the same helium abundance given that they have
nearly identical CMDs and ages; see V13.)  The turnoff luminosity is well
matched by a 12.9 Gyr isochrone for the same chemical abundances once the
predicted colors are adjusted by $-0.013$ mag in order to fit the observed TO
color. 

\begin{figure}[t]
\plotone{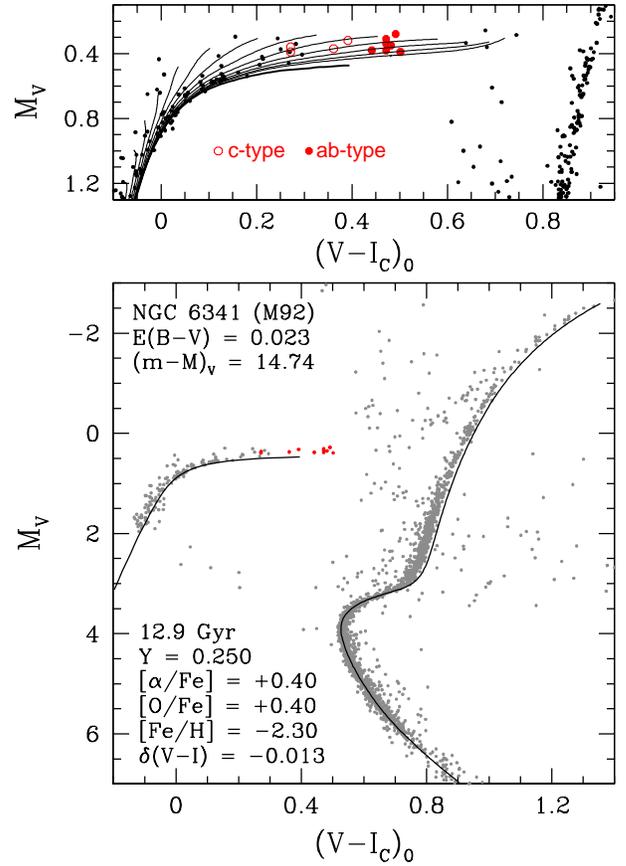}
\caption{{\it Bottom panel}: Overlay of an isochrone for the indicated age and
chemical abundances onto the CMD of M\,92 after the predicted colors have been
adjusted by $-0.013$ mag (as noted).  The adopted reddening and the apparent
distance modulus that has been derived from a fully consistent ZAHB, which has
been plotted without any adjustment to the predicted colors, is specified in
the top left-hand corner.  Member RR Lyrae are represented by small red dots.
{\it Top panel}: Superposition of the same ZAHB that appears in the bottom
panel, along with several evolutionary tracks for the core He-burning phase, 
onto the CMD for the HB and RGB populations of M\,92 that have $0.1 \le M_V \le
1.4$.  Only those tracks for models with ${\cal M} \le 0.71 \msol$\ have been
plotted.  Filled and open circles (in red) identify the $ab$-type and $c$-type
RR Lyrae stars, respectively.}
\end{figure}

The models faithfully reproduce the morphologies of the MS and RGB fiducial
sequences, though the predicted giant-branch location is too red by a few
hundredths of a magnitude.  Errors associated with the
adopted color--$\teff$\ relations, convection theory, the atmospheric boundary
conditions, or the assumed cluster parameters are some of the plausible
explanations for such discrepancies.  Note that the photometry was taken from
\citet[see their \S\,2]{vsb15}, who obtained a slightly older age
for M\,92 (13.0 Gyr), mainly because they adopted a lower [Fe/H] value by 0.1
dex.  Victoria models that assume higher values of [O/H] predict younger ages
(see, e.g., V13), which further highlights the sensitivity of absolute GC ages
to the adopted chemical abundances.

The same ZAHB that appears in the bottom panel of Fig.~9 is reproduced in the
top panel, where several tracks for the core He-burning phase are also plotted.
These follow the evolution of stars that arrive on the HB with the same
helium core mass --- but different envelope, and hence total, masses --- until
the central He abundance has decreased to $Y_C \sim 0.01$. (Because the tracks
for the more massive models follow nearly the same path towards the asymptotic
giant branch, making it very difficult to distinguish between them, only those
tracks for ${\cal M} \le 0.71 \msol$, which are the most relevant ones for the
interpretation of the cluster RR Lyrae, are shown.)  

The locations of the $ab$- and $c$-type RR Lyrae correspond to the values of
$\langle V\rangle_i$ and $(V-I_C)_m$ that were derived by
\citet{kop01}.  Unfortunately, it is not possible to improve upon these
estimates of their static equivalent colors because the necessary recipes are
not available: those given by \citet{bcs95}, which were used by \citet{ccc05}
for M\,3 variables, are restricted to the $B$, $V$, and $K$ bands only.  Based
on the differences between the values of $(B-V)_m$ and $(B-V)_S$
that are tabulated by Cacciari et al., one might expect that
$(V-I_C)_m$ colors should be corrected by about $-0.01$ mag in
order to better represent the colors of static stars.  Anyway, Fig.~9
shows that there is no color overlap of the fundamental and first-overtone
pulsators in M\,92.  In addition, it is apparent that the variables are all
significantly more luminous than the ZAHB at their colors and, judging from
the evolutionary sequences, they originate from ZAHB locations at $(V-I_C)_0
\lta 0.1$, where the majority of the non-variable HB stars are located.  Note
that the reddest ZAHB model, at $(V-I)_0 \approx 0.40$, is obtained if no mass
loss occurs during the preceding evolution.  To obtain redder ZAHB models with
[Fe/H] $= -2.30$, it is necessary to increase the assumed oxygen abundance (see
below).

\begin{figure}[t]
\plotone{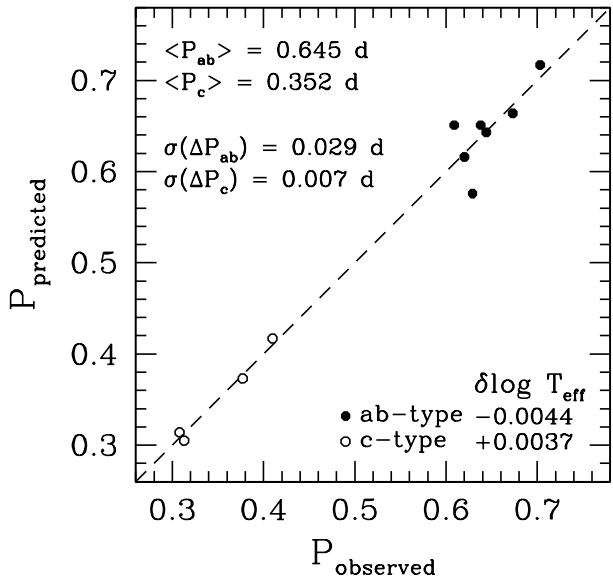}
\caption{Similar to Fig.~6, except that the observed periods of the RR Lyrae
in M\,92 are compared with those calculated using equations (1) and (2) on the
assumption of the luminosities, masses, and effective temperatures that are
obtained by interpolating within the grids of HB tracks at the CMD locations
of the cluster RR Lyrae stars (see the top panel in the previous figure).  The
$\delta\log\teff$\ offsets specified in the lower right-hand corner of the plot
were applied to the interpolated temperatures.}
\end{figure}

By interpolating within the grid of HB tracks, the values of $\log(L/L_\odot)$,
$\log({\cal M}/{\cal M}_\odot)$, and $\log\,\teff$ for each variable can be
derived, from which its period may be calculated using equation (1) or (2).
(For the models that appear in Fig.~9, $Z = 1.357\times 10^{-4}$.)  As discussed
in connection with Fig.~6, one can iterate on $\delta\log\teff$ adjustment that
is applied to the interpolated temperatures of the variables until the computed
values of \pab\ and \pc\ agree with the observed values.  The results obtained
via this procedure are illustrated in Figure~10.  The small dispersion in the
points about the dashed line, especially for the $c$-type variables, indicates
that the models do quite a good job of explaining the properties of the
RR Lyrae that reside in M\,92.  [If the two most discrepant $ab$-type
variables were removed from the sample, we would obtain $\sigma(\Delta P_{ab})
= 0.010$~d.  These stars are the bluest and the reddest filled circles in the
upper panel of Fig.~9 at $M_V \sim 0.38$.]

In fact, this conclusion is not strongly dependent on the assumption that
the colors of equivalent static stars correspond exactly to
$(V-I_C)_m$.  If these colors are adjusted by, e.g., $-0.01$ mag, one obtains a
virtually identical plot to that shown in Fig.~10 if the temperatures of 
the $ab$- and $c$-type variables are adjusted by $\delta\log\teff = -0.0071$ and
$+0.0013$, respectively.  These differences are still comparable to, or smaller
than the $1\sigma$ uncertainty in the model $\teff$\ scale.  [In making this
assertion, we are assuming that our models predict the temperatures of HB stars
just as well as in the case of turnoff stars at similar metallicities and
$\teff$ values.  For a discussion of the uncertainties in the temperatures of
main-sequence stars that are derived using the Infrared Flux Method (IRFM),
reference may be made to \citet{crm10}.  The success of modern isochrones in
matching the IRFM results to well within their uncertainties is demonstrated
by \citet{vcs10}.]
 
\begin{figure}[t]
\plotone{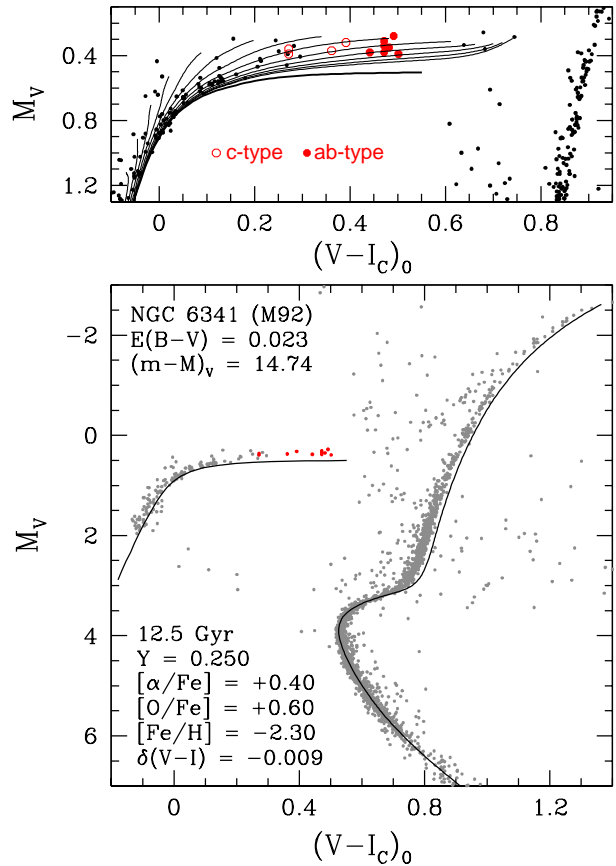}
\caption{As in Fig.~9, except that the stellar models assume [O/Fe] $= 0.6$.}
\end{figure}

A plot that is indistinguishable from Fig.~10 is also obtained if models for a
higher oxygen abundance by 0.2 dex (resulting in $Z = 1.786\times 10^{-4}$) 
are fitted to the observations (see Figure 11), provided that $-0.0021$ and
$+0.0053$ are adopted, in turn, for the $\delta\log\teff$ adjustments to the
temperatures predicted for the fundamental and first-overtone pulsators.  Higher
oxygen stretches metal-poor ZAHBs to redder colors and, at their red ends, to
slightly fainter $V$-band magnitudes (compare the ZAHBs for [O/Fe] $= 0.4$ and
0.6 in the upper panels of Figs.~9 and 11, respectively).  Both sets of models
assume the same abundances of helium and the other metals.  The main difference
between the tracks that pass through, or close to, the RR Lyrae in these plots
is a change in the predicted mass by $\approx 0.01\,\msol$.  For instance, the
track that intersects the reddest open circle in Fig.~9 was computed for a
$0.670\,\msol$ model, whereas the corresponding track in Fig.~11 assumed a mass
of $0.660\,\msol$.  The difference in mass is too small to have important
consequences for the predicted periods; as a result, Fig.~10 is relatively
insensitive to modest variations in [O/Fe].

Because the computed ZAHBs for [O/Fe] $=0.4$ and $0.6$ are nearly coincident at 
$(V-I_C)_0 \lta 0.05$, where the majority of the ``zero-age" HB stars in M\,92
appear to be located, essentially the same value of $(m-M)_V$ is implied by both
sequences.  However, turnoff luminosity versus age relations depend quite
strongly on the absolute abundance of oxygen (see, e.g., \citealt[their
Fig.~2]{vbn14}), or more generally [CNO/H] (assuming fixed solar abundances of
CNO).  Hence, as shown in the bottom panel of Fig.~11, the inferred
age of M\,92 is reduced by about 0.4 Gyr to 12.5 Gyr, if [O/Fe] $= 0.6$, as
compared with $\approx 12.9$ Gyr in Fig.~9, if the cluster stars have [O/Fe]
$= 0.4$.

A virtually identical fit to the MS, TO, and RGB populations of M\,92 can be
obtained from isochrones for [Fe/H] $ = -2.60$ and the same helium abundance
and metals mixture that are specified in Fig.~11 on the assumption of $E(B-V) =
0.023$ and $(m-M)_V = 14.78$ (as found from a fully consistent ZAHB).  The net
effect of assuming a lower value of [O/H] by 0.3 dex and a larger distance
modulus by 0.04 mag is to increase the predicted age to $\approx 12.8$ Gyr.
It turns out that the isochrone for this age reproduces the turnoff color
without requiring any adjustment of the predicted colors.  Aside from these
differences, it is not possible to distinguish between the fits of the [Fe/H]
$= -2.60$ and $-2.30$ isochrones to the turnoff and giant-branch photometry.
Accordingly, we have chosen to present the equivalent of just the top panel in
Fig.~11; i.e., a plot in which the ZAHB and selected HB tracks for [Fe/H] 
$= -2.60$ and [O/Fe] $= 0.6$ have been fitted to the cluster HB population.

As shown in Figure~12, the reddest ZAHB model has $(V-I_C)_0 \approx 0.25$,
which is considerably bluer than those plotted in Figs.~9 and 11 due to the
combined effects of lower [Fe/H] and (especially) reduced [O/H].  Nevertheless,
the superposition of the HB tracks onto the variable stars closely resembles
those shown previously.  In fact, the interpolated luminosities,
effective temperatures, and masses at the CMD locations of the RR Lyrae are
all sufficiently similar to those derived from the models for [Fe/H] $= -2.30$
that the periods calculated for them using equations (1) and (2), assuming the
appropriate value of $Z$ ($8.951\times 10^{-5}$), are not very different either.

To be more specific: the adoption of a larger distance modulus by 0.04 mag
implies higher luminosities by $\delta\log(L/L_\odot) \approx 0.016$ and higher
periods for the $ab$-type RR Lyrae by $\delta\log P_{ab} \approx 0.014$ (see
equation 1).  However, the predicted mass of each variable increases by
$\approx 0.015 \msol$ and the resultant changes to $\log\,P_{ab}$ given by the
$\log({\cal M}/\msol)$ and $\log\,Z$ terms in equation (1) amount to $\approx
-0.014$, with some minor star-to-star variations of these numbers.  (Basically
the same thing is found for the $c$-type variables.  Note that the predicted
temperatures at a given $V-I_C$ color do not change significantly if the [Fe/H]
value is reduced from $-2.30$ to $-2.60$.)  As a result, the comparison between
the predicted and observed periods can hardly be distinguished from that shown
in Fig.~10 if the inferred temperatures of the $ab$- and $c$-type pulsators
are adjusted by $\delta\log\teff = -0.0021$ and $+0.0063$, respectively.  As
before, these choices are set by the requirement that the models for [Fe/H]
$= -2.60$ and [O/Fe] $= 0.6$ predict the observed values of \pab\ and \pc.
Thus, Fig.~10 is obtained for M\,92 largely independently of moderate variations
in the adopted values of [Fe/H] and [O/Fe].

\begin{figure}[t]
\plotone{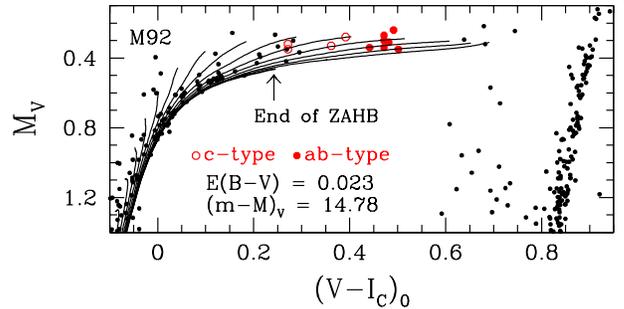}
\caption{Similar to the upper panel in the previous figure, except that the
ZAHB and evolutionary tracks (for ${\cal M} \le 0.72\msol$) that are compared
with the observations of M\,92 assume [Fe/H] $ = -2.60$.}
\end{figure}  
 
Although it is disappointing that the predicted periods of the RR Lyrae do not
provide a good constraint on the cluster metallicity, in view of the 
uncertainties associated with the former, it is nonetheless encouraging that
up-to-date HB models provide a satisfactory explanation of the properties of the
variable stars in both M\,92 and M\,3.  This includes, in particular, the
differences in \pab\ and \pc\ between them.  In addition, our findings support
the canonical understanding of the HB phase of evolution, given that the
faintest ``zero-age" cluster stars are matched exceedingly well by a ZAHB for
constant $Y$ over the entire range in color spanned by them.  Neither the fits
of ZAHB models to the cluster counterparts nor the comparisons between predicted
and observed RR Lyrae periods provide any compelling evidence for {\it large}
star-to-star helium abundance variations.  While the methods that we have
employed cannot detect the presence of modest variations (at the level of, say,
$\delta Y \lta 0.02$), any stars with $Y \gta 0.27$ that reside in M\,3 and/or
M\,92 must lie within the blue tails of their respective HB populations.   

One of the conclusions that can be drawn from the work described above is that
the distance moduli of M\,3 and M\,92 must be reasonably close to the values
implied by ZAHB models for $Y = 0.25$ and [Fe/H] values in the range of roughly
$-2.3$ to $-2.6$, as found spectroscopically.  (As shown in Figs.~9 and 11,
distances derived in this way are virtually independent of [O/Fe], which mainly
affects the predicted temperatures and colors of the more massive ZAHB models.
[Fe/H] uncertainties also have relatively minor ramifications for ZAHB-based
distance moduli given that $M_V$(HB) $\propto 0.21$\,[Fe/H] in the vicinity of
the instability strip; see V13, \citealt{cgb03}.)  Although our determination
of $(m-M)_V = 15.04$ for M\,3 agrees well with many estimates (e.g., 15.07 is
listed in the Harris catalog; see our footnote 4), the distance modulus of
M\,92 is more controversial.  Some discussion of this issue and of the
implications of our derived value of $(m-M)_V = 14.74$ for M\,92 is
warranted before we turn our attention to M\,15.  

\subsubsection{The distance modulus of M\,92}
\label{subsubsec:other}

Relatively short distance moduli have generally been derived for M\,92 when
nearby field halo subgiants, of which HD\,140283 is the most famous example,
are used as standard candles (e.g., \citealt{pmt98}, VandenBerg et al.~2002).
Such stars, which can be age-dated directly because they are located in the
region of a CMD where isochrones are most widely separated, are undeniably very
old.  The strongest evidence that they must have formed very soon after the
Big Bang is provided by the
work of \citet[also see \citealt{bnv13}]{vbn14}, who derived an age of
$14.3 \pm 0.8$ Gyr for HD\,140283 (where the stated uncertainty takes into
account all sources of error, including the parallax) using diffusive Victoria
models that were computed for metal abundances derived from high-resolution,
high S/N spectra.\footnote{A younger age by about 2\% would have been obtained
had FreeEOS, the sophisticated equation-of-state developed by A.~Irwin (see 
\S\ref{sec:models}), been used in this investigation.  Thus, the best estimate
of the age of HD\,140283 is closer to 14.0 Gyr than to 14.3 Gyr.  This is
still slightly older than the age of the universe from the analysis of
{\it Wilkinson Microwave Anisotropy Probe} observations ($13.77 \pm 0.06$ Gyr,
\citealt{blw13}), but the $1\,\sigma$ uncertainty associated with the stellar
age allows for the possibility that HD\,140283 formed within a few hundred Myr
after the Big Bang.}.

A few comments are in order concerning the latest study of HD\,140283 by
\citet{ctb15}, who found an age of $13.7 \pm 0.7$ Gyr (or less, if its
reddening is non-zero).  The somewhat younger age that they determined may be
due, in part, to their use of stellar evolutionary computations that, unlike
those employed by \citet{vbn14}, apparently did not take into account the 
important revisions to the rate of the $^{14}$N$(p,\,\gamma)^{15}$O reaction 
that occurred about a decade ago (\citealt[also see \citealt{mfg08}]{fic04}).
In addition, the low $\teff$ that Creevey et al.~derived for HD\,140283 can be
reproduced by stellar models only if very small values for the mixing-length
parameter ($\lta 1.0$) are assumed.  Such low values of $\amlt$\ have never been
found in studies of star cluster CMDs (see, e.g., \citealt{vsr00},
\citealt{scw02}), which provide far better constraints on the value of this
parameter than the properties of single stars (aside from the Sun) because the
location and slope of the giant branch, as well as the length of the SGB, are
very sensitive to the treatment of convection (see, e.g., \citealt{van83}).
These features cannot be reproduced unless a high value of $\amlt$\ is assumed
(see Figs.~2--4, 9, and 11 in the present paper).  In fact, 3D hydrodynamical
model atmospheres do not favor low values of $\amlt$\ either (\citealt{mwa15}).

Even though the solar-calibrated value of $\amlt$\ can vary significantly from
one study to the next, due to different assumptions concerning e.g., the
adopted solar abundances and the treatment of the surface boundary conditions,
isochrones for this value of the mixing-length parameter generally provide
credible fits to the CMDs of clusters for any metallicity.  This can be seen by
inspecting the plots provided by \citet{dcj08}, \citet{dvd12}, and
V14, whose models were computed on the assumption of solar-calibrated
values of $\amlt = 1.938$, 1.74, and 2.007, respectively.  The uncertainties of
the various factors that play a role in comparisons of isochrones with observed
CMDs are such that small variations in the mixing-length parameter with mass,
chemical abundances, or evolutionary state cannot be ruled out, but neither has
it been possible to argue compellingly in support of such variations (also see
\citealt{fvs06}).  Granted, there are indications from 3D model atmospheres that
$\amlt$\ {\it should} vary with $\teff$, gravity, and metallicity (e.g.,
\citealt{ts11}; \citealt[2015]{mca13}), but the first attempts to implement the
predictions from such simulations into stellar models have found that the
resultant tracks are not very different from those that assume constant $\amlt$
(\citealt{sc15}).

\citet{ctb15} have commented that the oxygen abundance that was derived by
\citet{vbn14} is based on a higher $\teff$\ than their determination.  However,
that abundance, [O/Fe] $=0.64$, agrees very well with the trends between [O/Fe]
and [Fe/H] given by \citet{dkb15} and \citet{aac15} for field Population II
stars that have [Fe/H] $< -2.0$.  To reduce the predicted age of HD\,140283, an
even higher O abundance would be required, which would suggest that the $\teff$
value adopted by VandenBerg et al.~should be {\it increased}.  That is, a cooler
$\teff$\ and the consequent decrease in [O/H] that would be needed to explain
the observed line strengths would tend to increase the discrepancy between the
age of the field subgiant and the age of the universe.  It is also worth
pointing out that a hot $\teff$\ scale is supported by the recent calibration
of the Infrared Flux Method by \citet{crm10}, the color-temperature relations
implied by MARCS model atmospheres (see CV14), and comparisons of
stellar models with the properties of solar neighborhood subdwarfs with
well-determined distances (\citealt[2014a]{vcs10}; \citealt{bsv10}).  The
spectroscopically derived temperature of HD\,140283 reported by \citet{vbn14}
is consistent with these photometric and theoretical determinations, but not
the one obtained by Creevey et al.  Further work is clearly needed to resolve
this controversy; in particular, a further examination of the
{\it model-dependent} aspects of the analysis of interferometric and
spectroscopic data carried out by Creevey et al.~may shed some light on
this difficulty.

Returning to the matter at hand: since metal-poor GCs are generally thought to
be among the oldest objects in the universe, one would expect that M\,92 and
HD\,140283, which appear to have very similar metallicities, would be nearly
coeval.  In this case, cluster subgiants with the same intrinsic color as
HD\,140283 should have the same absolute $V$-magnitude.  As shown in
Figure 13, this would imply that M\,92 has $(m-M)_V =
14.54$, which causes the cluster HB stars to be fainter than ZAHB models for
$Y = 0.25$, [Fe/H] $= -2.3$, [$\alpha$/Fe] $= 0.4$, and [O/Fe] $= 0.6$.  (These
abundances are close to those derived spectroscopically for HD\,140283; see
\citealt{vbn14}.)  This is less than the ZAHB-based distance modulus (see
Fig.~11) by 0.20 mag.

\begin{figure}[t]
\plotone{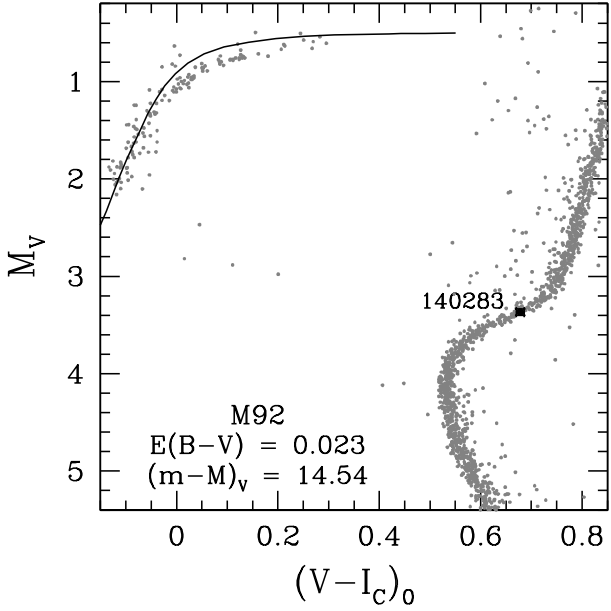}
\caption{Determination of the apparent distance modulus of M\,92 if its
SGB is matched to the CMD location of the field subgiant HD\,140283. 
This assumes that both objects are chemically indistinguishable and that they
have the same age, but it leads to irreconcilable differences between the ZAHB
models for [Fe/H] $= -2.30$ and the observed HB of M\,92.}
\end{figure}

However, we have already demonstrated that HB tracks for [Fe/H] $= -2.30$ are
able to explain the periods of the RR Lyrae in M\,92 quite well if the cluster
has $(m-M)_V = 14.74$ (see Fig.~9).  This would not be possible if 
the short distance modulus is assumed.  If the same tracks that are plotted in
the top panel of Fig.~11 were displaced to fainter magnitudes by 0.20 mag, which
corresponds to $\delta\log(L/L_\odot) \approx -0.08$, the predicted values of
\pab\ and \pc\ would decrease by $> 0.062$~d~according to equations (1) and
(2).  (The only way of explaining such a large offset is by assuming a helium
abundance that is much smaller than the primordial value of $Y$, which is not
justifiable.)  This provides a strong argument against such a faint HB, and we
therefore conclude that M\,92 subgiants of the same $(V-I_C)_0$\ color as
HD\,140283 must be intrinsically brighter than the field subgiant.  If
they are coeval and they have similar [Fe/H] values, M\,92 must have lower
[CNO/H] by $\gta 0.5$ dex --- but this is not supported by current spectroscopy;
e.g., see \citet{spk00}.  The most likely explanation is that M\,92 is younger
than HD\,140283 (by up to $\sim 1$ Gyr, depending primarily on the difference in
their respective CNO abundances).

Curiously, field Pop.~II subdwarfs seem to favor a larger distance modulus for
M\,92 than HD\,140283.  \citet{cfb13} have reported preliminary results for
three stars for which they obtained improved parallaxes using the {\it HST}
Fine Guidance Sensors.  Only one of them has [Fe/H] $< -2.0$; namely,
HD\,106924, which has [Fe/H] $= -2.13$, [O/Fe] $= 0.60$, $M_{F606W} = 5.96$
(with an uncertainty of about $\pm 0.015$ mag), and $m_{F606W}-m_{F814W} \approx
0.601$.  (These photometric properties were obtained by interpolating in their
Fig.~1.)  As shown in Figure~14, there is very little separation between
isochrones for [Fe/H] $= -2.0$ and $-2.6$ at the location of HD\,106924 on the
$[(m_{F606W}-m_{F814W})_0,\,M_{F606W}]$-diagram.  As a result, uncertainties in
the measured metallicity of the subdwarf should have no more than relatively
minor consequences.  (We note, however, that Chaboyer et al.~adopted a
significantly cooler $\teff$ for it than that implied by the MARCS color
transformations, so a metallicity $> -2.0$ cannot be entirely ruled out.)

\begin{figure}[t]
\plotone{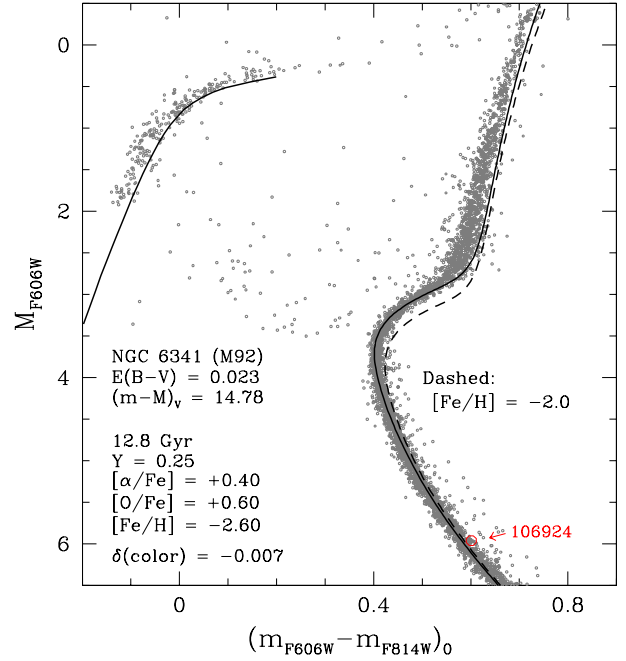}
\caption{Comparison of the location of the field subdwarf HD\,106924 (large red
open circle) with the CMD of M\,92 on the assumption of the indicated reddening
and ZAHB-based distance modulus.  The solid and dashed curves represent, in
turn, isochrones for [Fe/H] $= -2.6$ and $-2.0$ for the specified values of age,
$Y$, [$\alpha$/Fe], and [O/Fe].  The isochrones were adjusted to the blue by
0.007 mag.} 
\end{figure}

If M\,92 is assumed to have [Fe/H] $= -2.6$ (\citealt{rs11}) and the other
chemical abundance parameters have the indicated values, the ZAHB-based distance
modulus is $(m-M)_V = 14.78$ if $E(B-V) = 0.023$.\footnote{\citet[see their 
Fig.~7]{vbn14} noted that the nearby field giant HD\,122563 is redder than
M\,92 giants at the same $M_V$ by $\delta (B-V)_0 \approx 0.10$ mag.  This is
difficult to understand if M\,92 is more metal rich than HD\,122563, which has
[Fe/H] $\lta -2.6$ according to most spectroscopic studies (e.g.,
\citealt{cds04}, \citealt{rcl10}, \citealt{mgs11}).  Reasonable consistency of
the CMD locations of M\,92 giants and HD\,122563, implying a common metallicity,
would be obtained if the $M_V$ of HD\,122563 were adjusted by an amount
that corresponds to the $2\,\sigma$ parallax error bar.  While this paper was
being drafted, an article appeared by \citet{asf16}, who derived [Fe/H] $\sim
-2.9$ and $-2.7$ for HD\,122563 and HD\,140283, respectively, from IR spectra.
If the metallicity of HD\,140283 determined by \citet{vbn14} should be reduced
by 0.3 dex, their estimate of [O/Fe] should be increased by 0.3 dex to $\sim
0.95$ in order for the age of the subgiant to be compatible with the age of the
universe.  Such a high value of [O/Fe] seems inconsistent with most findings for
field halo stars that have similar metallicities (see \citealt{dkb15},
\citealt{aac15}).}  When these values are
adopted, HD\,106924 lies just to the red of the mean fiducial sequence of M\,92
at the observed subdwarf magnitude --- or, alternatively, HD\,106924 is slightly
brighter than cluster main-sequence stars that have the same color.  A better
centering of HD\,106924 onto the CMD of M\,92 would be obtained if
$(m-M)_V \approx 14.84$.  The uncertainties associated with the reddening and
the fit of the ZAHB to the cluster HB population certainly permit a larger
distance modulus by a few hundredths of a magnitude.  It is also possible that
the slight color offset of HD\,106924 relative to the M\,92 main sequence is
due to small zero-point differences in the photometry of the two
objects.

Another way of eliminating the apparent discrepancy is to adopt a higher He
abundance by $\delta Y \sim 0.015$, which implies a brighter HB, and thereby
an increased ZAHB-based distance modulus, by about 0.06 mag (or
$\delta\log(L/L_\odot) \approx 0.024)$.  We have checked that a ZAHB for
$Y = 0.265$ provides an equally good fit to the lower bound of the distribution
of HB stars in M\,92 as one for $Y = 0.25$ (see Figs.~9--12) when the 
aforementioned adjustment to the value of $(m-M)_V$ is adopted.  That is, such
a small change in $Y$ does not have detectable consequences for the quality of
the model fits to the observed CMD (including fits of isochrones to the
turnoff photometry). 

On the other hand, making the RR Lyrae stars brighter through
the adoption of a larger distance modulus would increase the predicted periods
of the variables; see equations (1) and (2).  However, the temperature
uncertainties are large enough that one could recover the results shown in
Fig.~10, on the assumption of $Y = 0.265$ instead of $Y = 0.25$, if higher
temperatures by only $\delta\log\teff \sim 0.006$ were adopted.  Since this is
within the $1\sigma$ error bar of the model $\teff$ scale, we conclude that
RR Lyrae periods alone cannot be used to provide a compelling argument in
support of a particular He abundance within the range $0.25 \lta Y \lta 0.265$.
Accurate distances based on, e.g., the best available calibration of the RR
Lyrae standard candle, which agree well with ZAHB-based distance determinations
(as described in \S~\ref{subsec:m3}), are needed to constrain the luminosities
of such variables.

The main conclusion to be drawn from Fig.~14 is that there is reasonably good
consistency between the distance modulus based on HD\,106924 and that derived
from ZAHB models.  In fact, this was the reason why we opted to use the
computations for [Fe/H] $= -2.6$ in this comparison instead of those for [Fe/H]
$= -2.3$, since a higher metallicity implies a smaller ZAHB-based distance
modulus by $\approx 0.04$ mag (see Figs.~9, 11).  However, this is admittedly a
weak argument in support of the possibility that M\,92 has [Fe/H] $\lta -2.6$
and [O/Fe] $= +0.6$.  A potential difficulty with these abundances is that, if
M\,92 and M\,15 have very similar chemical compositions, as is generally
believed to be the case, the ZAHB plotted in Fig.~14
is too blue to explain the large number
of RR Lyrae in M\,15 (recall our discussion in \S~\ref{sec:intro}).  In order
for that ZAHB to pass through the instability strip (as in the case of a ZAHB
for [Fe/H] $= -2.3$ and [O/Fe] $= +0.6$; see Fig.~11), a higher oxygen abundance
by $\gta 0.3$ dex would be needed, thereby resulting in [O/H] $\sim -1.7$ for
both [Fe/H] values.

\subsection{M\,15}
\label{subsec:m15}

As in the case of M\,92, most spectroscopic studies have found [Fe/H] $\approx
-2.3$ for M\,15 (\citealt{sjk00}, \citealt{ki03}, \citealt{cbs05},
\citealt{cbg09a}), but some of the same investigators now appear to favor 
values $\lta -2.6$ (\citealt{pst06}, \citealt{sks11}).  Because ZAHBs and core
He-burning tracks are much more dependent on [O/H] (and $Y$) than [Fe/H], a
0.3 dex reduction in the metallicity is not expected to have major consequences
for the interpretation of the M\,15 CMD {\it provided} that this change is
accompanied by a 0.3 dex increase in [O/Fe] (as obtained if the [O/H] value is
unchanged).  This may, in fact, be problematic for M\,15 since, as shown below,
it appears to be necessary to adopt [O/Fe] $\gta 0.8$, if [Fe/H] $= -2.3$ to
explain its RR Lyrae stars.  (Note that [O/Fe] values closer to $+0.3$ were
typically derived in spectroscopic studies of this GC in the 1990s; see, e.g.,
\citealt{sks97}.)  Consequently, models for [Fe/H] $= -2.6$ would yield a
similar interpretation of the data only if [O/Fe] $\gta 1.1$.  Because this
seems uncomfortably high (e.g., field stars of the same [Fe/H] typically
have [O/Fe] $\sim 0.75$; e.g., \citealt{dkb15}), we have decided to restrict
the present analysis to models for [Fe/H] $= -2.3$.  ([O/Fe] $\approx 0.8$ at
[Fe/H] $= -2.3$ is also on the high side, but [N/Fe] $\sim 1.6$ in some M\,15
giants (see \citealt{cbs05}) implies an initial O abundance corresponding to 
[O/Fe] $\sim 0.8$ if C$+$N$+$O is conserved and the same giants still have
[O/Fe] $\sim 0.3$ and [C/Fe] $< -0.5$.)

Turning to the photometry of M\,15: in their extensive study of the
$F606W,\,F814W$ observations of 55 GCs from the \citet{sbc07} survey, V13 found
that Victoria-Regina isochrones generally had to be shifted to the blue by
0.01--0.025 mag to match the observed turnoff color when reddenings from
\citet{sfd98}, metallicities from \citet{cbg09a}, and ZAHB-based distance
moduli were adopted.  In the case of
clusters with [Fe/H] $< -2.2$, M\,15 was the sole exception to this ``rule" in
that the requisite blueward shift was 0.038 mag, as compared with, e.g., 0.018
mag for M\,92 and 0.020 mag for M\,30.  According to Carretta et al., all three
of these clusters have the same [Fe/H] to within 0.02 dex, and V13 found that
they have the same age (12.75 Gyr).  Why, then, does M\,15 apparently have an
intrinsically bluer turnoff than M\,92 and M\,30?  [Interestingly, V13 (see
their Fig.~14) found that NGC\,2808 similarly stood out among GCs with $-1.0 >$
[Fe/H] $\ge -1.5$, and they suggested that isochrones for $Y = 0.25$ may require
an unusually large blueward color correction to match its turnoff because the
NGC\,2808 appears to contain stars with a wide range in helium abundance
(perhaps up to $Y = 0.40$; see \citealt{pba07}).  Is it possible that helium
abundance variations are significantly larger in M\,15 than in other GCs of
similar metallicity?]

To try to answer these questions, we will attempt to explain the properties of
the RR Lyrae variables that have been identified in M\,15 (though it is
expected that the highest-$Y$ stars would have very blue ZAHB locations and 
thus may not produce RR Lyrae stars).
However, let us first revisit the V13 analysis in the light of some improvements
that can be made to the CMDs of M\,15 and M\,92 and the use of different ZAHB
models and isochrones.  

Especially well-defined CMDs can be obtained by (i)
separating the MS and RGB stars in the \citet{sbc07} catalog from those that lie
to the blue of the giant branch, (ii) sorting the two samples into 0.1 mag bins,
and (iii) ranking the stars in each bin in terms of $\sigma_*$, where $\sigma_*$
is the smaller of the tabulated values of $\sigma_{F606W}$ and
$\sigma_{(F606W-F814W)}$.  If all stars with $\sigma_* > 0.02$ mag are excluded
and the remaining stars with the smallest photometric uncertainties, up to a
maximum number of 75 from each bin, are plotted, we obtain the CMDs for M\,15
and M\,92 that are shown in Figure~15.  The adopted selection procedure has
maximized the number of HB stars while limiting the vast number of MS (and RGB)
stars to those with the best photometry.  It is quite obvious from Fig.~15 that
the M\,15 CMD is considerably broader than that of M\,92 at any magnitude
(which may be due in part to the effects of differential reddening; see
\citealt{lbb15}.)

\begin{figure}[t]
\plotone{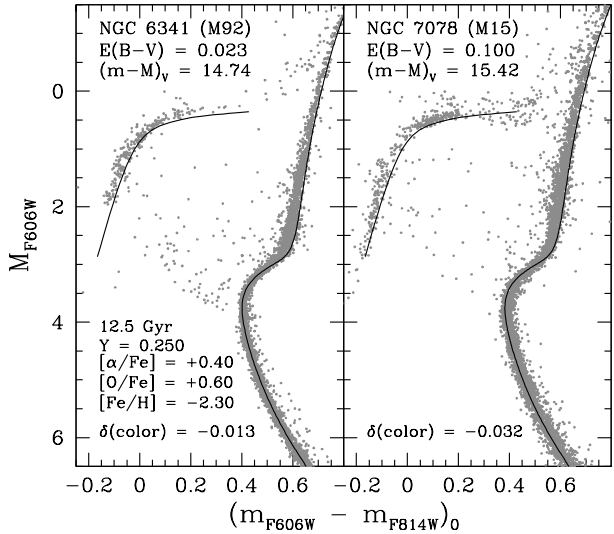}
\caption{{\it Left-hand panel}: As in the bottom panel of Fig.~11, except that
the models for the indicated chemical abundances and age are compared with
{\it HST} photometry of M\,92 (\citealt{sbc07}) rather than ground-based $VI_C$
data.  {\it Right-hand panel}: Fit of the same ZAHB and isochrone to {\it HST}
photometry of M\,15 (also from Sarajedini et al.).  Stars lying below the flat
part of the ZAHB and many of those above the densest concentration of HB stars
are RR Lyrae variables that have been observed at random phases of their
pulsation cycles.  Note that a different color offset had to be applied to the
isochrone to fit the turnoff colors of the two clusters (see the text for some
discussion of this point).}
\end{figure}

If the ZAHB and best-fit isochrone that appear in the bottom panel of Fig.~11
are transformed to $F606W,\,F814W$ magnitudes and compared with M\,92 photometry
on the assumption of the same reddening and apparent distance modulus, we
obtain the plot shown in the left-hand panel of Fig.~15.  To within the fitting
uncertainties, the same age ($\approx 12.5$ Gyr) is found on both color planes.
The right-hand panel shows that the same ZAHB provides a very good fit to the
HB population of M\,15 if $E(B-V) \approx 0.100$, which agrees very well with
recent determinations of line-of-sight reddenings from dust maps
(\citealt{sfd98}, \citealt{sf11}), and $(m-M)_V = 15.42$.  Under these
assumptions, the same age is found for M\,15 as for M\,92, though the best-fit
isochrone must be shifted by 0.032 mag to the blue to match the observed turnoff
color, as compared with 0.013 mag in the case of M\,92.

\begin{figure}[t]
\plotone{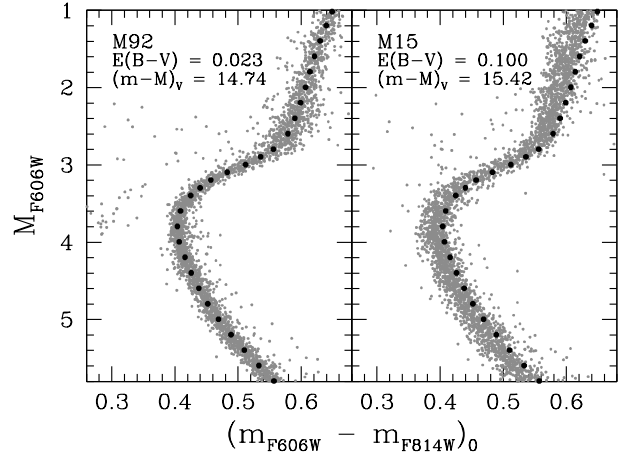}
\caption{{\it Left-hand panel}: Plot of the same M\,92 photometry that appears
in the previous figure for just the region of the CMD from the upper main
sequence to the lower RGB, along with the mean fiducial sequence that has been
derived from these stars (black filled circles).  The adopted values of $E(B-V)$
and $(m-M)_V$ are the same as those indicated in Fig.~15.  {\it Right-hand
panel}: As in the left-hand panel, except that the median M\,92 fiducial is
superimposed on the M\,15 CMD.}
\end{figure}

Figure~16 provides an alternative way of illustrating this difference.  The
left-hand panel reproduces the M\,92 photometry from the previous figure for
just the upper main sequence, subgiant, and lower RGB stars, on the assumption
of exactly the same reddening and distance modulus.  Using the methods described
by V13 (see their \S~5.3.1), the median locus through these stars was
determined; this is the sequence consisting of black filled circles that has
been superimposed on the smaller gray cluster stars.  When compared with the
M\,15 observations from Fig~15 for $5.8 \le M_{F606W} \le 1.0$ (see the
right-hand panel), this sequence is obviously too red by about 0.02 mag to
represent the M\,15 CMD.

One might conclude from this intercomparison that the adopted reddening of M\,15
is too high, since a much better superposition of the M\,15 and M\,92 turnoffs
would be obtained if M\,15 has $E(B-V) \approx 0.08$ rather than 0.10.  However,
as shown in Figure~17, such a low reddening presents problems for the
interpretation of the bluest HB stars in this cluster, as most of them would
then lie on the red side of the ZAHB at $M_{F606W} \gta 1.8$. (According to
canonical stellar evolutionary theory, the tracks of core He-burning stars
always remain brighter than the associated ZAHB locus at a given color.)  It
would therefore appear to be the case that M\,15 has an intrinsically bluer
turnoff than M\,92.  (Differential reddening in M\,15 could be partially
responsible for the apparent offset in the turnoff colors, but it is
unlikely to be the entire explanation; see below.)

\begin{figure}[t]
\plotone{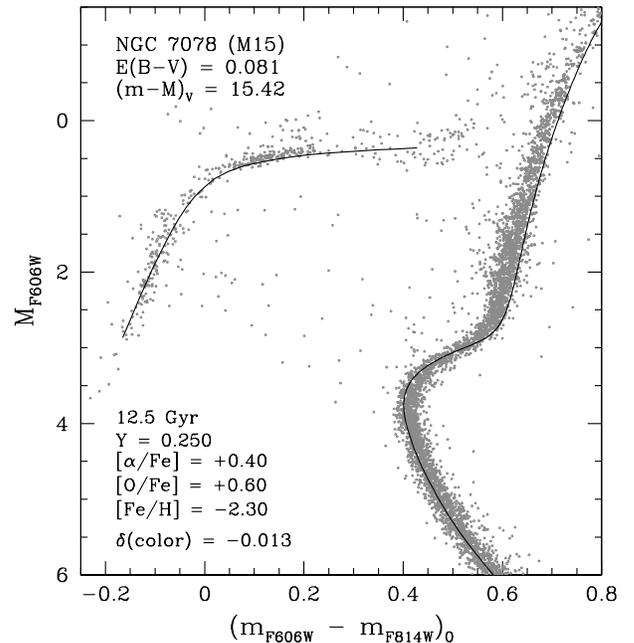}
\caption{As in the right-hand panel of Fig.~15, except that $E(B-V) = 0.081$
has been assumed so that M\,15 has the same intrinsic turnoff color as M\,92.
However, this leads to obvious problems with the fit of a ZAHB to the bluest
HB stars.}
\end{figure}

A difference in [Fe/H] (for which there is little support, anyway) would have
no more than a slight impact on the relative turnoff colors of M\,92 and M\,15
because the location of the MS on the [$(F606W-F814W)_0,\,M_{F606W}$]-plane has
almost no dependence on metallicity at [Fe/H] $\lta -2.3$ (as in the case of the
similar [$(V-I_C)_0,\,M_V$]-diagram; see \citealt{vcs10}).  Both clusters also
seem to have quite similar abundances of most of the so-called
$\alpha$-elements, such as Ca and Si (\citealt{spk00}).  We therefore suspect
that helium abundance differences are responsible; in particular, that
significantly larger {\it variations} in $Y$ are found in M\,15 than in M\,92.
An examination of the properties of the RR Lyrae in M\,15 should shed some light
on this possiblity.

\subsubsection{The RR Lyrae Stars in M\,15}
\label{subsubsec:m15rr}

Even though the \citet{cmd01} on-line catalog (see our footnote 9) has updated
information on the variable stars in M\,15 as recently as September 2014, it
is acknowledged therein that the most modern study of the cluster RR Lyrae is
still the one by \citet{cbs08}.   The main advance that has been made since
then is some clarification of variable identifications.  Using the astrometric
catalogs given by \citet{skp09}, Clement et al.~found that a few of the new
variables that Corwin et al.~claim to have discovered were, in fact, previously
known.  Since we are using the Corwin et al.~photometry (their Table 3) in the
present study, we have ensured that such misidentifications do not affect the
mean magnitudes and colors of the sample of variables that we have selected. 

Stars for which the authors could not measure reliable $B$-, $V$-, or
$I_C$-magnitudes were not considered, given the likelihood that the
mean magnitudes for those stars would not be very trustworthy, along with a
few stars that were either obviously too red (well outside the instability
strip) or too faint (significantly fainter than ZAHB loci).  Our final sample
consists of 56 RR Lyrae (29 $ab$-type, 27 $c$-type) that are presumed to have
reliable $B$ and $V$ measurements.  Of these stars, 38 (23 $ab$-type, 15
$c$-type) appear to have reliable $V$ and $I_C$ magnitudes.  

As in the case of M\,3 and M\,92, we checked that there is good consistency of
the interpretations of the [$(B-V)_0,\,M_V$]- and [$(V-I_C)_0,\,M_V$]-diagrams
for M\,15 with that shown in the right-hand panel of Fig.~15.  That is, the same
age is obtained, when the same distance modulus and reddening are assumed,
irrespective of whether the ZAHBs and isochrones are fitted to $BVI_C$ or
{\it HST} observations.  (The $BVI_C$ data were taken from the publicly
available ``Photometric Standard Fields" archive made available by
P.~Stetson; see, e.g.,
\citealt{st00}.\footnote{www.cadc.hia.nrc-cnrc.gc.ca/en/community/STETSON\hfill\break
/standards/}
Although there are relatively few blue HB stars in this dataset, there is a
sufficient number to show that $E(B-V) \approx 0.10$ is supported by the fit of
ZAHB models.)  The only important difference between the fits to the $BV$ and
$VI_C$ observations is that the ZAHB does not extend to sufficiently red $B-V$
colors to be fully consistent with the predictions of the same ZAHB on the
[$(V-I_C)_0,\,M_V$]-plane.

\begin{figure}[t]
\plotone{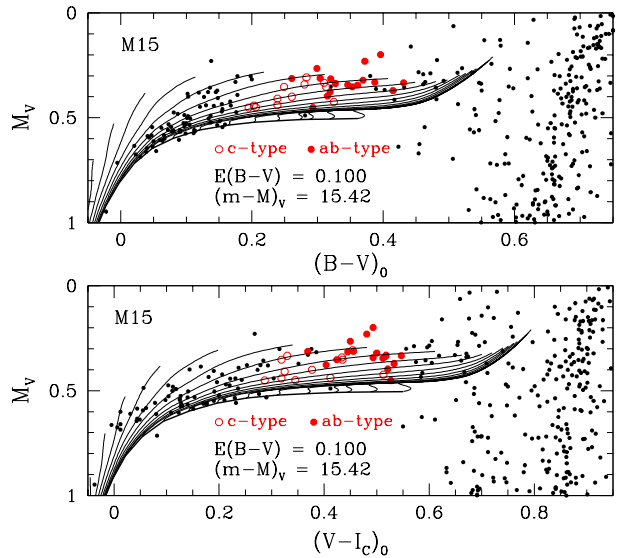}
\caption{Overlay of evolutionary tracks for core He-burning stars and the
corresponding ZAHB onto the CMD for the HB and RGB populations of M\,15 that
have $0.0 \le M_V \le 1.0$.  The same RR Lyrae stars appear in both panels.  The
ZAHB (for [Fe/H] $= -2.3$ and [O/Fe] $= 0.6$) is identical to the one that is
compared with {\it HST} photometry of M\,15 in the right-hand panel of Fig.~15,
and with $VI_C$ observations of M\,92 in Fig.~11.}  
\end{figure}

\begin{figure}[ht]
\plotone{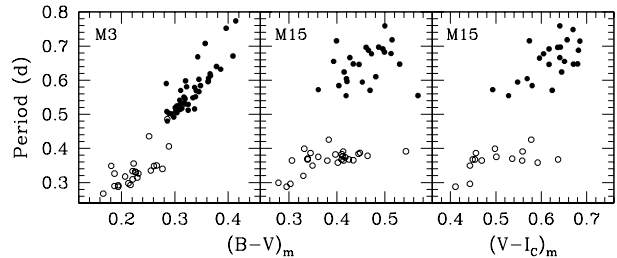}
\caption{Plot of the measured periods of the $ab$- and $c$-type variables 
(filled and open circles, respectively) in M\,3 as a function of $(B-V)_m$
(left-hand panel) and in M\,15 as a function of both $(B-V)_m$
(middle panel) and $(V-I_C)_m$ (left-hand panel).}
\end{figure}

This inconsistency is especially apparent if the ZAHB loci and associated
evolutionary tracks are compared with the locations of the M\,15 RR Lyrae,
assuming intensity-weighted $\langle V\rangle_i$ magnitudes and
magnitude-weighted $(B-V)_m$ or $(V-I_C)_m$ colors for the variables.
The upper panel of Figure~18 shows that many of the RR Lyrae
have redder $(B-V)_0$ colors than the reddest ZAHB model, whereas all of the 
variables have bluer $(V-I_C)_0$ colors than the reddest ZAHB model (see the
lower panel).  (Note that the same $ab$- and $c$-type variables are considered
on both color planes.)  Part of the explanation of this discrepancy could be
that $(B-V)_m$ needs a larger blueward correction than 
$(V-I_C)_m$ to represent the corresponding color of a static star,
though this should not amount to more than $\sim 0.02$ mag, judging from the
results for M\,3 by \citet{ccc05}.  Errors in the adopted $(B-V)$--$\teff$
transformations (from CV14) could also be a contributing factor.
Alternatively, errors in the derived values of the mean magnitudes and/or colors
may be primarily responsible for this conundrum.

In fact, the star-to-star scatter in the M\,15 data is much larger than in the
case of M\,3.  As illustrated in Figure~19, the periods of the variables in
M\,3 show a much tighter correlation with $(B-V)_m$ than those
in M\,15.  Particularly disconcerting is the fact that many of the
$c$-type variables in M\,15 span a very wide range in color despite having
nearly the same pulsation periods; these are the stars with $P \sim 0.38$~d.
Interestingly, the variables in M\,68, which appears to have the same [Fe/H] 
as M\,15 to within 0.1 dex (\citealt{cbg09a}), also has a remarkable
concentration of variables at a fixed period (see \citealt{cat04}, who provides
some discusson of this anomaly in both clusters).
Although helium abundance variations may provide a partial explanation,
this is unlikely to be the primary explanation because the properties of the
hotter first-overtone pulsators are considerably less dependent on $Y$ than the
cooler fundamental-mode pulsators.  This will become clear in the following
discussion.

Because $B-V$ colors will generally be more problematic than $V-I_C$, and
because we had satisfactory success explaining the properties of the RR Lyrae
in M\,92 when their effective temperatures were derived from $V-I_C$ colors, we
decided to present just our analysis of $VI_C$ photometry of M\,15.  Our main
goal, anyway, is to try to understand why the observed values of \pab\ are so
similar for M\,92 and M\,15, and this can be accomplished in a more robust way
if the same color is used to derive the temperatures of their respective
variable stars.  (The brief examination that we did carry out of the $B-V$ data
for M\,15 suggests that the measured $(B-V)_m$ values must be 
reduced by $\gta 0.03$ mag to obtain good consistency with the results
presented below.  Such offsets could be due, in part, to errors in the adopted
color--$\teff$ relations.)

\begin{figure}[t]
\plotone{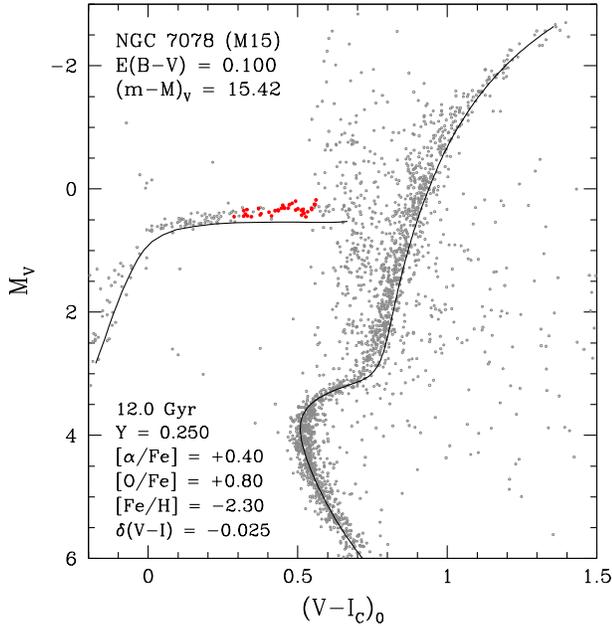}
\caption{As in the right-hand panel of Fig.~15, except that models for [O/Fe]
$= 0.8$ (instead of 0.6) have been fitted to the observed CMD of M\,15.
RR Lyrae stars have been plotted as small red dots.}
\end{figure}

It turns out that a ZAHB for [Fe/H] $= -2.3$ and [O/Fe] $= 0.6$ (i.e., for an
extra 0.2 dex above the amount implied by [$\alpha$/Fe] $= 0.4$) does not extend
far enough to the red in order for HB evolutionary tracks to explain the RR
Lyrae if some/most of them have $Y > 0.25$, so we opted to use models for
[O/Fe] $= 0.8$ in our analysis.  This does not affect the ZAHB-based distance
modulus, but it does imply a reduced turnoff age by $\approx 0.5$ Gyr.  As
shown in Figure~20, the best-fit isochrone for the higher oxygen abundance
predicts an age of $\approx 12.0$ Gyr for M\,15 if it has $(m-M)_V = 15.42$ and
$E(B-V) = 0.10$.  Under these assumptions, a ZAHB for $Y = 0.25$, [Fe/H]
$= -2.3$, and [O/Fe] $= 0.8$ (with [m/Fe] $= 0.4$ for the other
$\alpha$-elements) provides quite a good fit to the faintest cluster HB stars.
(Recall from \S~\ref{subsubsec:modm92} that a 0.2 dex increase in [O/Fe] has
only minor consequences for the predicted periods of RR Lyrae variables.)

Possible fits of ZAHB loci and the associated evolutionary tracks for the core
He-burning phase are illustrated in Figure~21.  The bottom panel shows that a
ZAHB for $Y = 0.25$ provides a good fit to the non-variable stars just to the
blue of the instability strip and that all of the RR Lyrae are located well
above this ZAHB.  Indeed, many of the $ab$-type variables are located near, or
past, the ends of tracks, where relatively few stars are expected because of the 
increasingly rapid rate of evolution as the central helium content is depleted
(e.g., \citealt{psc02}).  (Recall that the tracks end when $Y_C \approx 0.01$.)
As discussed in \S~\ref{sec:intro}, the large number of variables has always
been a strong argument that most of them cannot be highly evolved stars, but
rather that the majority must be relatively close to their respective ZAHB
locations.  (Note that the selected stars represent only $\sim 25$--30\% of the
total number of RR Lyrae in M\,15, which is especially rich in these variables;
see \citealt{cbs08}.)

\begin{figure}[t]
\plotone{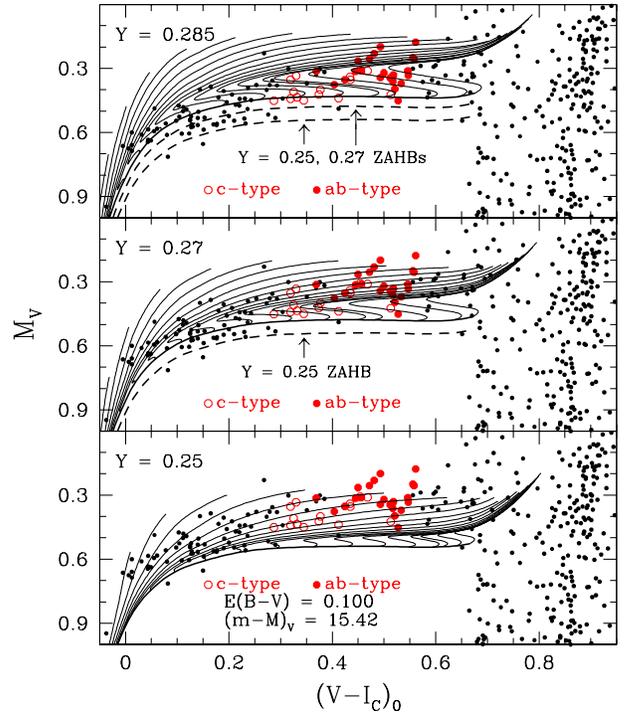}
\caption{Overlays of ZAHBs and evolutionary tracks for different helium
abundances onto the HB population of M\,15.  Filled and open circles (in red)
identify the $ab$- and $c$-type RR Lyrae, respectively.  The dashed loci in
the middle and upper panels reproduce the ZAHBs that appear in lower panels 
(for lower $Y$, as indicated).  The reddening and distance modulus that are
specified in the lower panel apply to all three panels.}
\end{figure}

Higher $Y$ is one way of achieving this since, as shown in the middle and upper
panels of Fig.~21, many (or most) of the variables would be located along, or
just above, ZAHBs for $Y \gta 0.27$.  This, together with the increased
prominence of blue loops in tracks for a given mass and higher helium
abundance, means that the masses of the RR Lyrae stars that are derived by
interpolating within the evolutionary tracks will increase with increasing $Y$.
(That the effect on the mass can be quite large for some of the stars is obvious
from an inspection of Fig.~21.)  Indeed, it is mainly through the mass terms in
equations (1) and (2) that the predicted periods will be affected by the
variable's helium abundance.  If all of the stars are at the same distance,
subject to the same reddening, and have the essentially same metal abundances,
the contributions to $\log\,P$ arising from the $\log(L/L_\odot)$,
$\log\,\teff$, and $\log Z$ terms will be nearly, or entirely, independent
of $Y$. 

The fact that the coefficients of the mass terms are relatively small in
equations (1) and (2) makes it difficult to use the observed periods to obtain a
clear separation of the M\,15 variables into groups of normal (i.e., close to
primordial), intermediate, and high helium abundances.  At best, predicted
periods will be uncertain by $\delta P \approx \pm 0.03$~d, which corresponds
(for instance) to an error bar of $\sim 0.01$ dex in the $\log\teff$ value of an
individual RR Lyrae.  Consequently, the models for $Y = 0.25$, 0.27, and 0.285
are likely to fare equally well, or equally poorly, in explaining the observed
periods --- though one can anticipate that, due to vagaries in the data, one
helium abundance might be favored over the others depending on whether the
predicted period for that $Y$ is just inside or just outside its assumed
$1\,\sigma$ uncertainty. 

\begin{figure}[t]
\plotone{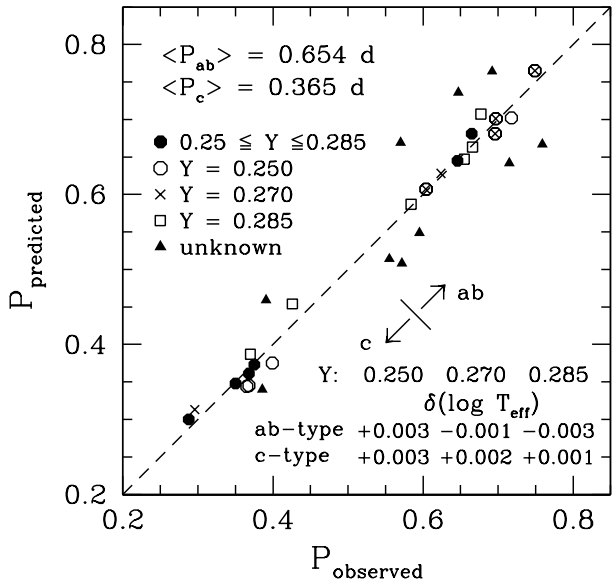}
\caption{Similar to Fig.~6 (for M\,3) and Fig.~10 (for M\,92), except that the
M\,15 variables are considered.  Symbols identify RR Lyrae with periods that
are reproduced to within $\pm 0.03$~d~by models for $Y = 0.25$ (open circles),
0.27 (crosses), or 0.285 (open squares).  Filled circles are used if the models
for all three values of $Y$ satisfy this criterion, while filled triangles
represent stars for which the discrepancies between the predicted and observed
periods are $> 0.03$~d.  The $\delta\log\teff$\ offsets that were applied to 
the models for $Y = 0.25$, 0.27, and 0.285 so that the predicted values of
\pab\ and \pc\ agree with the observed values are given in the lower right-hand
corner. The differences between the predicted and observed periods have
$\sigma(\Delta P_{ab}) = 0.050$~d and $\sigma(\Delta P_c) = 0.030$~d.  Averages
of the predicted periods and the stellar properties have been adopted for those
variables that are plotted as filled circles, the superposition of two different
symbols, or filled triangles.} 
\end{figure}

These remarks anticipate the results of our interpolations, which are presented
in Figure~22.  This shows that our models for {\it any} of the three values of
$Y$ that we have considered are able to reproduce the observed pulsation periods
of 6 RR Lyrae (the ones represented by filled circles) to within $\pm
0.03$~d~(our adopted consistency criterion).  Not surprisingly, most of them
are $c$-type variables, for which the range in the interpolated masses
from the grids of HB tracks for $Y = 0.25$, 0.27, and 0.285 is typically $\lta
0.05 \msol$, as compared with a range that can be as large as $\sim 0.1 \msol$
for many of the cooler $ab$-type variables.  The periods of a few other
variables can be explained quite well by models for $Y = 0.25$ and 0.27, which
are identified by the superposition of open circles and crosses, while
a preference for just one of the three values of $Y$ is obtained for several
other stars (those plotted as open circles, open squares, or crosses).
Satisfactory explanations of the measured periods could not be obtained for 
those stars that are represented by filled triangles.  

There were a few variables in the initial sample for which the discrepancies
between the predicted and observed periods were even larger than those obtained
for any of the stars plotted in Fig.~22.  They were dropped from consideration
prior to performing the analysis that produced the results shown in this figure,
because their removal made it much easier to achieve consistency between the
predicted and observed values of \pab\ and \pc.  As indicated in Fig.~22, the
$\delta\log\teff$ adjustments that were applied to the RR Lyrae temperatures to
attain this consistency are all quite small.  There is no reason to retain the
extreme outliers in the sample anyway, as it is not possible to explain, in
the context of the adopted stellar models, why the periods
implied by their CMD locations would be so much larger or smaller than the
observed periods (by $\sim 0.10$--0.22~d) when reasonably good agreement is
found for the majority of the variables.  (The particularly problematic RR Lyrae
stars are the fundamental pulsators V21, V47, and V60, and the first-overtone 
pulsators V70 and V120.  Further study of these variables is clearly needed to
to understand their apparently anomalous properties.)

In order to make our analysis of the M\,15 variables as close to a purely
differential study with respect to M\,92 as possible, we fitted the same $Y =
0.25$, [Fe/H] $= -2.3$, [O/Fe] $= 0.8$ models that we have employed for M\,15 to
observations of M\,92.  By interpolating within the corresponding grids of HB
tracks, we found that the observed values of \pab\ and \pc\ for M\,92 could
be reproduced if the temperatures of the $ab$- and $c$-type pulsators were
adjusted by $\delta\log\teff = 0.0$ and $+0.0069$, respectively.  These
determinations differ by $\sim 0.004$ from the adjustments adopted in the 
construction of Fig.~10, which is essentially identical with the plot that is
obtained on the assumption of models that assume [O/Fe] $= 0.8$.  Small
shifts in the RR Lyrae $\teff$\ scale that are needed to reconcile the predicted
and observed values of \pab\ and \pc\ in M\,15, on the one hand, and in M\,92,
on the other, are plausibly due to differences in the photometric zero-points
and/or in the respective magnitude-weighted colors.

Although our analysis suggests that many of the RR Lyrae in M\,15 have a
higher helium abundance than those in M\,92, the large number of variables in
M\,15 provides more compelling support for the same conclusion.  It is widely
accepted that, as in the case of M\,3, ``the bulk of the RR Lyrae
in M\,15 are in an early stage of evolution from the ZAHB" (\citealt{bcd84}).
However, the ZAHB that fits the nonvariable blue HB stars
in M\,15, which provides an equally good fit to their counterparts in M\,92, is
considerably fainter than the lowest-luminosity RR Lyrae (in both clusters).
Whereas the small number of variables in M\,92 is consistent with them being
evolved stars from ZAHB structures well to the blue of the instability strip,
the large number of variables in M\,15, which have similar or greater
luminosities than those residing in M\,92, argues that there must be at least
two distinct ZAHB populations in M\,15.  The most likely explanation for the
offset in luminosity between them is a difference in the abundance of helium ---
a conclusion that is supported by our detailed examination of the cluster RR
Lyrae stars.  [The extended blue tail of M\,15's HB, which contains a
non-uniform distribution of stars with gaps at some magnitudes, has long been
thought to indicate the presence of multiple stellar populations (see, e.g.,
\citealt{bcf85}, \citealt{cro88}).  These observations may be indicative of
even larger helium abundance variations than the $\delta Y \sim 0.03$--04 that
is probably sufficient to account for the stars along the flat part of its HB.]

\section{Summary and Discussion}
\label{sec:sum}

\begin{table*}[t]
\begin{center}
\caption{Mean Properties of the M\,3, M\,15, and M\,92 RR Lyrae Stars}
\begin{tabular}{rcccccccccccccc}
\hline
\hline
Name & & \multicolumn{1}{c}{\per\tablenotemark{a}} & $\sigma$ & \multicolumn{1}{c}{\per\tablenotemark{b}}
     & $\sigma$ & \tavg & $\sigma$ & \mavg & $\sigma$ & \bavg & $\sigma$ & \vavg & $\sigma$ & $Z$ \\ [0.5ex]
\hline
\multicolumn{15}{c}{$ab$-type} \\ [0.5ex]
M\,3  & & 0.568 & 0.067 & 0.568 & 0.075 & 3.812 & 0.012 & 0.656 & 0.016 & 0.534 & 0.058 & 0.583 & 0.056 & $7.623\times 10^{-4}$ \\ 
M\,15 & & 0.654 & 0.060 & 0.654 & 0.068 & 3.813 & 0.013 & 0.706 & 0.051 & 0.257 & 0.055 & 0.326 & 0.058 & $2.466\times 10^{-4}$ \\
M\,92 & & 0.645 & 0.033 & 0.645 & 0.042 & 3.815 & 0.005 & 0.673 & 0.011 & 0.283 & 0.042 & 0.347 & 0.041 & $1.786\times 10^{-4}$ \\ [0.5ex]
\multicolumn{15}{c}{$c$-type} \\ [0.5ex]
M\,3  & & 0.336 & 0.050 & 0.336 & 0.058 & 3.855 & 0.013 & 0.630 & 0.016 & 0.514 & 0.104 & 0.520 & 0.102 & $7.623\times 10^{-4}$ \\
M\,15 & & 0.365 & 0.038 & 0.365 & 0.046 & 3.848 & 0.014 & 0.712 & 0.038 & 0.363 & 0.060 & 0.395 & 0.052 & $2.466\times 10^{-4}$ \\ 
M\,92 & & 0.352 & 0.050 & 0.352 & 0.053 & 3.860 & 0.016 & 0.662 & 0.005 & 0.340 & 0.040 & 0.360 & 0.029 & $1.786\times 10^{-4}$ \\ [0.5ex]
\hline
\multicolumn{15}{l}{$^{\rm a}$Observed mean period (in d) of the samples of RR Lyrae considered in this study.} \\
\multicolumn{15}{l}{$^{\rm b}$Predicted mean period (in d) of the samples of RR Lyrae considered in this study.} \\
\end{tabular}
\end{center}
\end{table*}

Mainly during the last 3 decades of the 20$^{\rm th}$ century, but continuing to
the present day, many investigators in the GC, stellar evolution, and variable
star communities have tried to understand the \citet[1944]{oo39} dichotomy,
particularly as regards M\,3 (Oo type I) and M\,15 (Oo II) because they are so
rich in RR Lyrae.  No one worked harder to explain the difference in the mean
periods of their RR Lyrae populations than Allan Sandage and, in the end, it
seems that his solution to this problem, that M\,15 RR Lyrae stars have higher
helium abundances than those in M\,3 (see \citealt{sks81}), stands a good chance
of being the right answer.  [Prior to $\sim 2005$, GCs were considered to be
simple stellar populations in which all stars within each cluster were thought
to be coeval and essentially chemically homogeneous, aside from the ubiquitous
star-to-star variations in CN.  It was generally assumed that helium did not
vary, given that the application of the R-method (\citealt{ib68}) yielded very
similar helium abundances, $Y \approx 0.25$, for most clusters, especially those
with red HBs (e.g., see \citealt[and references therein]{src04}).  Consequently,
everyone viewed the possibility that $Y$ varies inversely with [Fe/H], which
also seems counter-intuitive from a chemical evolution perspective, with
considerable skepticism.  Only recently has it been established that GCs contain
multiple, chemically distinct stellar populations that have, or probably have,
different helium abundances.  As mentioned in \S~\ref{sec:intro}, the most
massive clusters have provided the most compelling evidence for such variations.]

To be sure, our apparent success in modeling the HBs of M\,3, M\,15, and M\,92
is due in part to the improvements made to both the stellar models over the
years and the theoretical relations that describe the dependence of the period
on luminosity, mass, $\teff$, and metallicity for the fundamental and the
first-overtone pulsators (\citealt{mcb15}).  With just minor adjustments
(well within the associated uncertainties) to the RR Lyrae $\teff$\ scale (or,
equivalently, to the coefficients of $\log\teff$\ in these equations), it is
possible to reproduce the observed values of \pab\ and \pc\ quite well.  The
advances that have been made likely explain why we find $\delta Y$(M\,15
{\it minus} M\,3) $\sim 0.03$, as compared with a difference closer to 0.05, in
the same sense, that was derived by \citet[also see \citealt{srt87}]{sks81}.

Table~1 lists the observed and predicted values of \pab\ and \pc\ and their
standard deviations ($\sigma$), as derived from the individual RR Lyrae
stars in each cluster, along with the mean values and standard deviations of
the temperatures, masses, absolute bolometric magnitudes, and absolute $V$-band
magnitudes of the variables.  (Note that the mean periods which are calculated
from equations (1) and (2) on the assumption of the quantities given in the
sixth, eighth, tenth, and last columns agree very well with the values listed
in the fourth column.)  Nearly the same value of \tavg\ is obtained for
the $ab$-type RR Lyrae ($\approx 3.815$) and $c$-type variables ($\approx 
3.855$) in all three clusters. 

In addition, the variables in M\,15 are 
predicted to have higher mean masses (and significantly larger mass dispersions)
than those in M\,92 and M\,3.  Consistent with the plots of the RR Lyrae on
various CMDs (see, e.g., Figs.~5, 12, and 18), the magnitudes of the
first-overtone pulsators in M\,3 have the largest standard deviations, while
the luminosity dispersions of both the $ab$- and $c$-type RR Lyrae are the
smallest in M\,92.  Not surprisingly, the variables in M\,15 and M\,92 have
brighter absolute magnitudes by $\gta 0.2$ mag than those in M\,3.  (As one
would expect, the tabulated properties have some dependence on the samples of
RR Lyrae that are considered.  For instance, had we retained the most
problematic M\,15 variables in our analysis, we would have obtained
\bavg\ $= 0.241 \pm 0.068$ and \vavg\ $= 0.313 \pm 0.065$ for the $ab$-type
variables in this cluster, as well as \bavg\ $= 0.354 \pm 0.062$ and
\vavg\ $= 0.392 \pm 0.053$ for its first-overtone pulsators.) 

The importance of taking diffusive processes into account should be appreciated.
One of the consequences of diffusion is that, due to the settling of helium
in the interiors of stars during their main sequence and subgiant evolution,
the envelope helium abundance after the first dredge-up (i.e., after the
convective envelope has reached its maximum depth on the lower RGB) is predicted
to be {\it less} than the initial helium content by $\delta Y \sim 0.003$
(assuming a $0.8 \msol$ model for [Fe/H] $= 1.55$).  If diffusion is ignored,
the envelope helium abundance is predicted to be {\it higher} than the initial
abundance by $\delta Y \sim 0.01$.  Since the luminosity of the HB is a
sensitive function of $Y$ (see Figs.~8, 21), ZAHBs will be significantly
fainter, implying reduced ZAHB-based distance moduli, if diffusion is treated.
(As discussed in \S~\ref{subsec:m3}, the value of $(m-M)_V = 15.04$ that
we have derived for M\,3 using ZAHB models satisfies the constraint provided
by current calibrations of the RR Lyrae standard candle quite well.)

Furthermore, the prominence of blue loops in post-ZAHB evolutionary tracks
depends quite strongly on the helium abundance (see, e.g., Fig.~21).  This
has important ramifications for the intermingling of $ab$-type and $c$-type
RR Lyrae.  For instance, as discussed in \S~\ref{subsec:m3}, there appears to
be very little overlap of the colors of these variables in M\,3 (see Fig.~5),
which suggests that blue loops must be small or non-existent if the
{\it transition} between fundamental and first-overtone pulsation, or vice
versa, depends on the direction in which the core He-burning stars are evolving
through the instability strip (the so-called ``hysteresis effect"; see
\citealt[and especially the very instructive plots provided by \citealt{cct78}
and \citealt{san81}]{vb73}).  Our diffusive models for an initial helium
abundance of $Y = 0.250$ predict small blue loops (see Fig.~5), though better
consistency with the observations would be obtained if they were even smaller,
which suggests that a slightly lower value of $Y$ should be adopted (but within
the uncertainties of the primordial helium abundance) or that our models
underestimate the rate at which settling occurs in stars.  

Higher $Y$ by $\sim 0.013$, as predicted by non-diffusive stellar models, would
result in extended blue loops, which seems incompatible with the fairly sharp
boundary between the $ab$- and $c$-type variables in M\,3.  (The age of
HD\,140283 provides another argument that diffusion physics should not be
neglected in stellar models; see \S~\ref{subsubsec:other} and \citealt{vbn14}.)
There is no overlap of the colors of these RR Lyrae in M\,92, nor is any
expected because the evolution through the instability strip is clearly in the
direction from blue to red from ZAHB locations on the blue side of the
instability strip (see Figs.~9, 11, 12).  It is not clear what to make
of the M\,15 variables in this regard (see Fig.~21), partly because the
magnitude-weighted colors derived by \citet{cbs08} seem particularly uncertain,
and because it is not possible to unambiguously determine the helium
abundances of the individual RR Lyrae stars.  Further observational work 
to determine improved estimates of the colors of equivalent static stars of the
M\,15 variables would be very helpful, as would an extension, towards lower
metallicities and additional bandpasses, of the theoretical studies of
\citet{bcs95} on the differences between different types of averages and the
static magnitudes and colors.

As discussed by \citet{amb15}, a separation of the $ab$- and $c$-type
RR Lyrae in terms of their colors is found in most GCs, irrespective of
whether they are OoI or OoII systems.  Notable exceptions are the OoI cluster
NGC\,3201 (\citealt{aac14}) and the OoII clusters M\,15 and $\omega$\ Cen (see
\citealt{san81}).  When there is a mixture of fundamental and first-overtone
pulsators in a restricted color range within the instability strip, some of the
variables are expected to be evolving from red to blue, and their periods
should be decreasing with time, while others will be evolving in the opposite
direction with positive period-change rates, $\Delta P/\Delta t$ (see Fig.~3
by \citealt{cct78}). 

In principle, it should be possible to use measurements of
$\Delta P/\Delta t$ to determine the directions in which individual variables
are evolving.  In practice, however, this seems to be very difficult.  As
\citet{cc01} have concluded, period-change rates in M\,3 appear to be due more
to ``noise" than to evolutionary effects.
In their follow-up of the Corwin \& Carney study, \citet{jhs12} noted that 
``positive and negative period-change rates with similar size are equally
frequent at any period and brightness".  For instance, V1 and V10 have
comparable mean magnitudes and colors, but the values of $\Delta P/\Delta t$
tabulated for them by Jurcsik et al.~are $-0.417$~d/Myr and $+0.343$~d/Myr,
respectively.  These stars should both have increasing periods judging from
their CMD locations, which are well above the ZAHB, relative to our evolutionary
sequences.  The same can be said of even brighter variables, and yet, as
reported by Jurcsik et al., none of the four brightest RR Lyrae in M\,3 have 
$\Delta P/\Delta t > 0.0$.  This includes the brightest $ab$-type variable in
our sample, V42, which has a period-change rate of $-1.132$~d/Myr.  In view of
such results, and the possible concerns with the mean magnitudes and colors of
M\,15 variables mentioned above, we have not attempted to pursue this line of
investigation --- though it may be worthwhile to do so when improved data
become available.

Being able to explain the RR Lyrae in M\,3 and M\,92 so well provides valuable
support for our determinations of their distance moduli and ages.  We find no
compelling evidence for helium abundance variations in either cluster from our
analysis of the variable stars, though star-to-star differences at the level of
$\delta Y \lta 0.02$ would be very difficult to detect.  Our analysis suggests
that the faintest HB stars on the blue side of the instability strip and some
of the RR Lyrae in M\,15 represent an M\,92-like population.  The fact that the 
difference in magnitude between these stars and the turnoff is identical to
within measuring uncertainties in both clusters leads us to conclude that M\,15
and M\,92 are coeval (as most previous studies have found).  However, M\,15
appears to contain additional populations of stars with
higher helium abundances, up to at least $Y \sim 0.29$ in the vicinity of the
instability strip and possibly to higher values along the extended blue tail of
the cluster HB.  (Fits of isochrones to the turnoff photometry on the assumption
of ZAHB-based distance moduli suggest that M\,3, M\,15, and M\,92 all have ages
of $\approx 12.5$ Gyr, depending on the assumed CNO abundances.  It seems
unlikely, in fact, that GCs are as old as the field halo subgiant HD\,140283
--- which is not implausible given the recent discovery of a galaxy at a
redshift $z = 11.1$ that seems to have built-up a stellar mass of $\sim 10^9
\msol$ within just $\sim 400$ Myr after the Big Bang; see \citealt{obv16}.) 

Our explanation of the M\,15 RR Lyrae does raise an important question: why
are the ZAHB stars with higher helium abundances distributed to redder colors
than those for $Y \approx 0.25$?  Possible answers to this question are (i)
mass loss rates vary inversely with $Y$, though we are unaware of any
empirical or theoretical support for this suggestion, (ii) the 
abundances of the CNO elements are higher in stars with increased helium
abundances, (iii) the stars with higher $Y$ are somewhat younger than those
with normal $Y$, or (iv) some combination of these possibilities.  \citet{jlj14}
have suggested that second generation stars would have enhanced helium and CNO
abundances though, in their proposed explanation of the Oosterhoff dichotomy,
the different generations of stars would span different color ranges.  We see
no evidence that this is the case; indeed, a ZAHB for $Y = 0.25$ appears to
provide a good fit to the faintest HB stars at all $(V-I_C)_0 \lta 0.3$, where
stars with higher helium are presumably also found.

More importantly, there is very little spectroscopic evidence for variations
in the total CNO abundance in M\,15.  Athough they studied only a few giants,
\citet{sks97} found that C$+$N$+$O is constant to within the measurement
uncertainties in 5 of the 6 stars in their sample.  One giant apparently has
much higher CNO, given that the derived nitrogen abundance corresponds to
[N/Fe] $\sim +1.6$.  However, previous studies of much larger samples of upper
RGB stars concluded that there are no real differences in the C and N abundances
of M\,15 and M\,92 (\citealt{crl82}, \citealt{tlc83}).  Both clusters do show a
steep decline of [C/Fe] with [Fe/H] (also see \citealt{bbs01}), but C--N and
O--N cycling together with deep mixing (\citealt{dd90}, \citealt{lhs93},
\citealt{dv03}) can explain those observations without requiring star-to-star
variations in CNO (\citealt{pi88}, \citealt{skp91}, \citealt{cbs05}).

Although C$+$N$+$O seems to be approximately constant, the observed variations
of CN in present-day main sequence and subgiant stars, as well as star-to-star
differences in Mg and Al at any luminosity (e.g., see the relevant studies of
M\,92 and M\,15 by \citealt{ksb98}, \citealt{gvb00}, \citealt{cbs05},
\citealt{cbg09b}), is an entirely different issue because they cannot be
produced by evolutionary processes within lower mass stars.  Such variations
must have arisen during an extended period (or possibly successive epochs) of
star formation at early times or, if the stars currently observed within a
given GC are coeval, they must have been present in the gas out of which the
cluster stars formed.  That the helium-enhanced stars in M\,15 appear to 
populate ZAHBs that extend to much redder colors than the ZAHB which fits the
faintest stars to the blue of the instability strip suggests that the spread in
stellar ages may be larger in M\,15 than in M\,92 or M\,3.  [Regardless of which
scenario provides the most correct explanation, there is little doubt that
H-burning nucleosynthesis at very high temperatures ($\sim 75 \times 10^6$~K,
as predicted for supermassive stars; see \citealt{dh14}) is responsible for the
observed abundance correlations and anti-correlations, including ratios of the
abundances of magnesium isotopes (\citealt{dvh15}).]

The potential importance of rotation for our understanding of the HBs in GCs
should be kept in mind as well.  In the few studies that have been undertaken 
during the past few decades to measure the rotation of member stars,
unexpectedly high rotational velocities have been determined for blue HB stars
in the most metal-poor systems (specifically, M\,92; see \citealt{cm97}) and in
higher metallicity GCs with extremely blue HBs, including M\,13 (\citealt{pe83})
and NGC\,288 (\citealt{pe85}).  A spread of rotational velocities in the
precursor red giants would seem to be the most probable cause of the variations
in total mass along GC horizontal branches, and the average mass loss could
therefore be related to the average rotational velocity in upper RGB stars.
The fact that the decline of the surface carbon abundance with increasing
luminosity along the giant branch is much more pronounced in M\,13
(\citealt{sm03}) and most, if not all, of the extremely metal-deficient GCs
(\citealt{msb08}) than in metal-rich clusters can hardly be a coincidence.
\citet{sm79} have shown that such observations can be explained by rotationally
driven deep-mixing, which is expected to become less important as the
metallicity increases due to the concomitant increase in the mean molecular
weight gradient near the H-burning shell.  (Red giants in NGC\,288, which has a
higher [Fe/H] than M\,13 by $\sim 0.5$ dex, mainly show a bimodality of CN
strengths with only a hint of a decline of C and O with increasing luminosity;
see \citealt{sl09}.)  

\smallskip
The next paper in the series will carefully examine the differences in the CMDs
of M\,3 and M\,13 to try to explain why these two clusters have such different
HB morphologies, despite having nearly identical [Fe/H] values.  Synthetic HB
populations will be presented in this investigation, which will include an
analysis of, among other things, the detailed distribution of mass along the
observed HBs of these two systems, which is an important and controversial
issue; see, e.g., \citet{cd08} and \citet{vc08}.

\acknowledgements

This paper has benefitted from a thoughtful report by an anonymous
referee, to whom we are very grateful.  Both D.A.V.~and M.C.~acknowledge with
deep appreciation the encouragement and inspiration provided by Allan Sandage
over the years.  Financial support from the Natural Sciences and Engineering
Research Council via a Discovery Grant to D.A.V.~is also acknowledged with
gratitude.  Support for M.C.~is provided by the Ministry for the Economy,
Development, and Tourism's Millennium Science Initiative through grant
IC\,120009, awarded to the Millennium Institute of Astrophysics (MAS); by
Proyecto Basal PFB-06/2007; and by FONDECYT grant \#1141141.

\end{document}